\documentclass[longauth]{aa}  
\usepackage{amsmath,amssymb,amsfonts}
\usepackage{graphicx}
\usepackage{txfonts}
\usepackage{enumerate} % advanced enumerate environment
\usepackage{colordvi} % for color text
\usepackage{bm}% bold math
\usepackage{epsfig}
\usepackage{txfonts}
\usepackage{natbib}
\usepackage{scalerel}
\usepackage[pdfencoding=auto,psdextra]{hyperref}
\hypersetup{
    colorlinks=true,
    linkcolor=blue,
    filecolor=magenta,      
    urlcolor=blue,
    citecolor=blue
}
\usepackage[dvipsnames, table]{xcolor}
\usepackage{color, colortbl}
\usepackage[T1]{fontenc} % if needed
\usepackage{braket}
\usepackage{euclid}
\usepackage{placeins}
\usepackage{tabularx}
\usepackage[switch, mathlines, modulo]{lineno}
\usepackage{ulem}
\let\oldequation\equation
\let\oldendequation\endequation

\renewenvironment{equation}
  {\linenomathNonumbers\oldequation}
  {\oldendequation\endlinenomath}
  
\let\oldalign\align
\let\oldendalign\endalign

\renewenvironment{align}
  {\linenomathNonumbers\oldalign}
  {\oldendalign\endlinenomath}

\definecolor{amethyst}{rgb}{0.6, 0.4, 0.8}

\makeatletter
\renewcommand*\aa@pageof{, page \thepage{} of \pageref*{LastPage}}
\makeatother

\definecolor{amaranth}{rgb}{0.9, 0.17, 0.31}
\definecolor{forestgreen(web)}{rgb}{0.13, 0.55, 0.13}
\definecolor{lavender(web)}{rgb}{0.9, 0.9, 0.98}
\definecolor{cosmiclatte}{rgb}{1.0, 0.97, 0.91}
\definecolor{jonquil}{rgb}{0.98, 0.85, 0.37}
\definecolor{khaki(x11)(lightkhaki)}{rgb}{0.94, 0.9, 0.55}
\definecolor{thistle}{rgb}{0.85, 0.75, 0.85}

\newcommand{\GCsp}{\text{GC}\ensuremath{_\mathrm{sp}}}
\newcommand{\GCph}{\text{GC}\ensuremath{_\mathrm{ph}}}
\newcommand{\XCph}{\text{XC}\ensuremath{_\mathrm{ph}}}
\newcommand{\Omegam}{\ensuremath{\Omega_{\mathrm{m},0}}}
\newcommand{\Omegab}{\ensuremath{\Omega_{\mathrm{b},0}}}
\newcommand{\OmegaDE}{\ensuremath{\Omega_{\mathrm{DE},0}}}

\newcommand{\lcdm}{\ensuremath{\Lambda\mathrm{CDM}}}
\newcommand{\logfr}{\ensuremath{\logten|f_{R0}|}}
\newcommand{\fr}{\ensuremath{|f_{R0}|}}

\newcommand{\de}{\mathrm{d}}

\defcitealias{Blanchard:2019oqi}{EC19}

\usepackage[nameinlink, capitalise]{cleveref}
\usepackage{orcidlink}

\crefname{chapter}{Chap.}{Chaps.}
\crefname{section}{Sect.}{Sects.}
\crefname{figure}{Fig.}{Figs.}
\Crefname{chapter}{Chapter}{Chapters}
\Crefname{section}{Section}{Sections}
\Crefname{figure}{Figure}{Figures}

\begin{document}
%\runningpagewiselinenumbers
%\linenumbers
\title{\Euclid: Constraints on f(R) cosmologies from the spectroscopic and photometric primary probes\thanks{This paper is published on behalf of the Euclid Consortium.}}

%% please do not edit the author list -- contact ECEB Bureau for changes
\newcommand{\orcid}[1]{\orcidlink{#1}}
\author{S.~Casas\orcid{0000-0002-4751-5138}$^{1}$\thanks{\email{s.casas@protonmail.com}}, V.~F.~Cardone$^{2,3}$, D.~Sapone\orcid{0000-0001-7089-4503}$^{4}$, N.~Frusciante$^{5}$, F.~Pace\orcid{0000-0001-8039-0480}$^{6,7,8}$, G.~Parimbelli\orcid{0000-0002-2539-2472}$^{9,10,11,12}$, M.~Archidiacono$^{13,14}$, K.~Koyama$^{15}$, I.~Tutusaus\orcid{0000-0002-3199-0399}$^{16,17,18,19}$, S.~Camera\orcid{0000-0003-3399-3574}$^{6,7,8}$
% ...
}

%%%% please do not edit the affiliation list -- contact ECEB Bureau for changes
\institute{$^{1}$ Institute for Theoretical Particle Physics and Cosmology (TTK), RWTH Aachen University, 52056 Aachen, Germany\\
$^{2}$ INAF-Osservatorio Astronomico di Roma, Via Frascati 33, 00078 Monteporzio Catone, Italy\\
$^{3}$ INFN-Sezione di Roma, Piazzale Aldo Moro, 2 - c/o Dipartimento di Fisica, Edificio G. Marconi, 00185 Roma, Italy\\
$^{4}$ Departamento de F\'isica, FCFM, Universidad de Chile, Blanco Encalada 2008, Santiago, Chile\\
$^{5}$ Department of Physics "E. Pancini", University Federico II, Via Cinthia 6, 80126, Napoli, Italy\\
$^{6}$ Dipartimento di Fisica, Universit\'a degli Studi di Torino, Via P. Giuria 1, 10125 Torino, Italy\\
$^{7}$ INFN-Sezione di Torino, Via P. Giuria 1, 10125 Torino, Italy\\
$^{8}$ INAF-Osservatorio Astrofisico di Torino, Via Osservatorio 20, 10025 Pino Torinese (TO), Italy\\
$^{9}$ Dipartimento di Fisica, Universit\'a degli studi di Genova, and INFN-Sezione di Genova, via Dodecaneso 33, 16146, Genova, Italy\\
$^{10}$ INAF-Osservatorio Astronomico di Trieste, Via G. B. Tiepolo 11, 34143 Trieste, Italy\\
$^{11}$ SISSA, International School for Advanced Studies, Via Bonomea 265, 34136 Trieste TS, Italy\\
$^{12}$ IFPU, Institute for Fundamental Physics of the Universe, via Beirut 2, 34151 Trieste, Italy\\
$^{13}$ Dipartimento di Fisica "Aldo Pontremoli", Universit\'a degli Studi di Milano, Via Celoria 16, 20133 Milano, Italy\\
$^{14}$ INFN-Sezione di Milano, Via Celoria 16, 20133 Milano, Italy\\
$^{15}$ Institute of Cosmology and Gravitation, University of Portsmouth, Portsmouth PO1 3FX, UK\\
$^{16}$ Institut de Recherche en Astrophysique et Plan\'etologie (IRAP), Universit\'e de Toulouse, CNRS, UPS, CNES, 14 Av. Edouard Belin, 31400 Toulouse, France\\
$^{17}$ Universit\'e de Gen\`eve, D\'epartement de Physique Th\'eorique and Centre for Astroparticle Physics, 24 quai Ernest-Ansermet, CH-1211 Gen\`eve 4, Switzerland\\
$^{18}$ Institute of Space Sciences (ICE, CSIC), Campus UAB, Carrer de Can Magrans, s/n, 08193 Barcelona, Spain\\
$^{19}$ Institut d'Estudis Espacials de Catalunya (IEEC), Carrer Gran Capit\'a 2-4, 08034 Barcelona, Spain\\
$^{20}$ Universit\'e Paris-Saclay, Universit\'e Paris Cit\'e, CEA, CNRS, Astrophysique, Instrumentation et Mod\'elisation Paris-Saclay, 91191 Gif-sur-Yvette, France\\
$^{21}$ Institut f\"ur Theoretische Physik, University of Heidelberg, Philosophenweg 16, 69120 Heidelberg, Germany\\
$^{22}$ Universit\'e St Joseph; Faculty of Sciences, Beirut, Lebanon\\
$^{23}$ Institute Lorentz, Leiden University, PO Box 9506, Leiden 2300 RA, The Netherlands\\
$^{24}$ Dipartimento di Scienze Matematiche, Fisiche e Informatiche, Universit\`a di Parma, Viale delle Scienze 7/A 43124 Parma, Italy\\
$^{25}$ INFN Gruppo Collegato di Parma, Viale delle Scienze 7/A 43124 Parma, Italy\\
$^{26}$ Institut de Physique Th\'eorique, CEA, CNRS, Universit\'e Paris-Saclay 91191 Gif-sur-Yvette Cedex, France\\
$^{27}$ Mullard Space Science Laboratory, University College London, Holmbury St Mary, Dorking, Surrey RH5 6NT, UK\\
$^{28}$ Institute for Astronomy, University of Edinburgh, Royal Observatory, Blackford Hill, Edinburgh EH9 3HJ, UK\\
$^{29}$ Higgs Centre for Theoretical Physics, School of Physics and Astronomy, The University of Edinburgh, Edinburgh EH9 3FD, UK\\
$^{30}$ Universit\'e Paris-Saclay, CNRS, Institut d'astrophysique spatiale, 91405, Orsay, France\\
$^{31}$ INAF-IASF Milano, Via Alfonso Corti 12, 20133 Milano, Italy\\
$^{32}$ Instituto de F\'isica Te\'orica UAM-CSIC, Campus de Cantoblanco, 28049 Madrid, Spain\\
$^{33}$ ESAC/ESA, Camino Bajo del Castillo, s/n., Urb. Villafranca del Castillo, 28692 Villanueva de la Ca\~nada, Madrid, Spain\\
$^{34}$ INAF-Osservatorio di Astrofisica e Scienza dello Spazio di Bologna, Via Piero Gobetti 93/3, 40129 Bologna, Italy\\
$^{35}$ Dipartimento di Fisica e Astronomia, Universit\'a di Bologna, Via Gobetti 93/2, 40129 Bologna, Italy\\
$^{36}$ INFN-Sezione di Bologna, Viale Berti Pichat 6/2, 40127 Bologna, Italy\\
$^{37}$ Max Planck Institute for Extraterrestrial Physics, Giessenbachstr. 1, 85748 Garching, Germany\\
$^{38}$ Dipartimento di Fisica, Universit\'a di Genova, Via Dodecaneso 33, 16146, Genova, Italy\\
$^{39}$ INFN-Sezione di Genova, Via Dodecaneso 33, 16146, Genova, Italy\\
$^{40}$ INAF-Osservatorio Astronomico di Capodimonte, Via Moiariello 16, 80131 Napoli, Italy\\
$^{41}$ Instituto de Astrof\'isica e Ci\^encias do Espa\c{c}o, Universidade do Porto, CAUP, Rua das Estrelas, PT4150-762 Porto, Portugal\\
$^{42}$ Institut de F\'{i}sica d'Altes Energies (IFAE), The Barcelona Institute of Science and Technology, Campus UAB, 08193 Bellaterra (Barcelona), Spain\\
$^{43}$ Port d'Informaci\'{o} Cient\'{i}fica, Campus UAB, C. Albareda s/n, 08193 Bellaterra (Barcelona), Spain\\
$^{44}$ INFN section of Naples, Via Cinthia 6, 80126, Napoli, Italy\\
$^{45}$ Dipartimento di Fisica e Astronomia "Augusto Righi" - Alma Mater Studiorum Universit\'a di Bologna, Viale Berti Pichat 6/2, 40127 Bologna, Italy\\
$^{46}$ Centre National d'Etudes Spatiales -- Centre spatial de Toulouse, 18 avenue Edouard Belin, 31401 Toulouse Cedex 9, France\\
$^{47}$ Institut national de physique nucl\'eaire et de physique des particules, 3 rue Michel-Ange, 75794 Paris C\'edex 16, France\\
$^{48}$ Jodrell Bank Centre for Astrophysics, Department of Physics and Astronomy, University of Manchester, Oxford Road, Manchester M13 9PL, UK\\
$^{49}$ European Space Agency/ESRIN, Largo Galileo Galilei 1, 00044 Frascati, Roma, Italy\\
$^{50}$ University of Lyon, Univ Claude Bernard Lyon 1, CNRS/IN2P3, IP2I Lyon, UMR 5822, 69622 Villeurbanne, France\\
$^{51}$ Institute of Physics, Laboratory of Astrophysics, Ecole Polytechnique F\'ed\'erale de Lausanne (EPFL), Observatoire de Sauverny, 1290 Versoix, Switzerland\\
$^{52}$ UCB Lyon 1, CNRS/IN2P3, IUF, IP2I Lyon, 4 rue Enrico Fermi, 69622 Villeurbanne, France\\
$^{53}$ Departamento de F\'isica, Faculdade de Ci\^encias, Universidade de Lisboa, Edif\'icio C8, Campo Grande, PT1749-016 Lisboa, Portugal\\
$^{54}$ Instituto de Astrof\'isica e Ci\^encias do Espa\c{c}o, Faculdade de Ci\^encias, Universidade de Lisboa, Campo Grande, 1749-016 Lisboa, Portugal\\
$^{55}$ Department of Astronomy, University of Geneva, ch. d'Ecogia 16, 1290 Versoix, Switzerland\\
$^{56}$ Department of Physics, Oxford University, Keble Road, Oxford OX1 3RH, UK\\
$^{57}$ INFN-Padova, Via Marzolo 8, 35131 Padova, Italy\\
$^{58}$ Universit\'e Paris-Saclay, Universit\'e Paris Cit\'e, CEA, CNRS, AIM, 91191, Gif-sur-Yvette, France\\
$^{59}$ Aix-Marseille Universit\'e, CNRS/IN2P3, CPPM, Marseille, France\\
$^{60}$ INAF-Osservatorio Astronomico di Padova, Via dell'Osservatorio 5, 35122 Padova, Italy\\
$^{61}$ University Observatory, Faculty of Physics, Ludwig-Maximilians-Universit{\"a}t, Scheinerstr. 1, 81679 Munich, Germany\\
$^{62}$ INAF-Osservatorio Astronomico di Brera, Via Brera 28, 20122 Milano, Italy\\
$^{63}$ Institute of Theoretical Astrophysics, University of Oslo, P.O. Box 1029 Blindern, 0315 Oslo, Norway\\
$^{64}$ von Hoerner \& Sulger GmbH, Schlo{\ss}Platz 8, 68723 Schwetzingen, Germany\\
$^{65}$ Technical University of Denmark, Elektrovej 327, 2800 Kgs. Lyngby, Denmark\\
$^{66}$ Cosmic Dawn Center (DAWN), Denmark\\
$^{67}$ Institut d'Astrophysique de Paris, UMR 7095, CNRS, and Sorbonne Universit\'e, 98 bis boulevard Arago, 75014 Paris, France\\
$^{68}$ Max-Planck-Institut f\"ur Astronomie, K\"onigstuhl 17, 69117 Heidelberg, Germany\\
$^{69}$ Jet Propulsion Laboratory, California Institute of Technology, 4800 Oak Grove Drive, Pasadena, CA, 91109, USA\\
$^{70}$ Department of Physics, P.O. Box 64, 00014 University of Helsinki, Finland\\
$^{71}$ Helsinki Institute of Physics, Gustaf H{\"a}llstr{\"o}min katu 2, University of Helsinki, Helsinki, Finland\\
$^{72}$ NOVA optical infrared instrumentation group at ASTRON, Oude Hoogeveensedijk 4, 7991PD, Dwingeloo, The Netherlands\\
$^{73}$ Argelander-Institut f\"ur Astronomie, Universit\"at Bonn, Auf dem H\"ugel 71, 53121 Bonn, Germany\\
$^{74}$ Dipartimento di Fisica e Astronomia "Augusto Righi" - Alma Mater Studiorum Universit\'a di Bologna, via Piero Gobetti 93/2, 40129 Bologna, Italy\\
$^{75}$ Department of Physics, Institute for Computational Cosmology, Durham University, South Road, DH1 3LE, UK\\
$^{76}$ CEA Saclay, DFR/IRFU, Service d'Astrophysique, Bat. 709, 91191 Gif-sur-Yvette, France\\
$^{77}$ European Space Agency/ESTEC, Keplerlaan 1, 2201 AZ Noordwijk, The Netherlands\\
$^{78}$ Department of Physics and Astronomy, University of Aarhus, Ny Munkegade 120, DK-8000 Aarhus C, Denmark\\
$^{79}$ Centre for Astrophysics, University of Waterloo, Waterloo, Ontario N2L 3G1, Canada\\
$^{80}$ Department of Physics and Astronomy, University of Waterloo, Waterloo, Ontario N2L 3G1, Canada\\
$^{81}$ Perimeter Institute for Theoretical Physics, Waterloo, Ontario N2L 2Y5, Canada\\
$^{82}$ Space Science Data Center, Italian Space Agency, via del Politecnico snc, 00133 Roma, Italy\\
$^{83}$ Institute of Space Science, Str. Atomistilor, nr. 409 M\u{a}gurele, Ilfov, 077125, Romania\\
$^{84}$ Dipartimento di Fisica e Astronomia "G. Galilei", Universit\'a di Padova, Via Marzolo 8, 35131 Padova, Italy\\
$^{85}$ Universit\"ats-Sternwarte M\"unchen, Fakult\"at f\"ur Physik, Ludwig-Maximilians-Universit\"at M\"unchen, Scheinerstrasse 1, 81679 M\"unchen, Germany\\
$^{86}$ Institut d'Astrophysique de Paris, 98bis Boulevard Arago, 75014, Paris, France\\
$^{87}$ Junia, EPA department, 41 Bd Vauban, 59800 Lille, France\\
$^{88}$ Satlantis, University Science Park, Sede Bld 48940, Leioa-Bilbao, Spain\\
$^{89}$ AIM, CEA, CNRS, Universit\'{e} Paris-Saclay, Universit\'{e} de Paris, 91191 Gif-sur-Yvette, France\\
$^{90}$ Aix-Marseille Universit\'e, CNRS, CNES, LAM, Marseille, France\\
$^{91}$ Centro de Investigaciones Energ\'eticas, Medioambientales y Tecnol\'ogicas (CIEMAT), Avenida Complutense 40, 28040 Madrid, Spain\\
$^{92}$ Instituto de Astrof\'isica e Ci\^encias do Espa\c{c}o, Faculdade de Ci\^encias, Universidade de Lisboa, Tapada da Ajuda, 1349-018 Lisboa, Portugal\\
$^{93}$ Universidad Polit\'ecnica de Cartagena, Departamento de Electr\'onica y Tecnolog\'ia de Computadoras,  Plaza del Hospital 1, 30202 Cartagena, Spain\\
$^{94}$ Kapteyn Astronomical Institute, University of Groningen, PO Box 800, 9700 AV Groningen, The Netherlands\\
$^{95}$ INFN-Bologna, Via Irnerio 46, 40126 Bologna, Italy\\
$^{96}$ Infrared Processing and Analysis Center, California Institute of Technology, Pasadena, CA 91125, USA}

\date{\today}

\authorrunning{Casas et al.}

\titlerunning{Euclid: forecasts on $f(R)$ cosmologies.}

% \abstract{}{}{}{}{} 
% 5 {} token are mandatory

\abstract
{\Euclid will provide a powerful compilation of data including spectroscopic redshifts, the angular clustering of galaxies, weak lensing cosmic shear, and the cross-correlation of these last two photometric observables. This will lead to very stringent constraints on the \lcdm\ concordance cosmological model and models beyond it, in which for instance standard gravity is modified.
}
{In this study we extend recently presented \Euclid forecasts into the Hu-Sawicki $f(R)$ cosmological model, a popular extension of the Hilbert-Einstein action that introduces an universal modified gravity force in a scale-dependent way. 
This scale-dependent modification requires a generalisation of our previous recipes for spectroscopic and photometric galaxy clustering, and for weak lensing, both in the linear and in the non-linear regimes.
Our aim is to estimate how well future \Euclid data will be able to constrain the extra parameter of the theory, $f_{R0}$, for the range in which this parameter is still allowed by current observations.}
{For the spectroscopic probe, we use a phenomenological approach to account for the scale dependence of the growth of perturbations in the terms related to baryon acoustic oscillations and redshift-space distortions. For the photometric observables, which probe deeper into the non-linear regime, we use a fitting formula developed in the literature that captures the modifications in the non-linear matter power spectrum caused by the $f(R)$ model.} 
{We show that, in an optimistic setting, and for a fiducial value of $\fr = 5 \times 10^{-6}$, \Euclid alone will be able to constrain the additional parameter $\logfr$ at the $3\%$ level, using spectroscopic galaxy clustering alone; 
at the $1.4\%$ level, using the combination of photometric probes on their own; 
and at the $1\%$ level, using the combination of spectroscopic and photometric observations. 
This last constraint corresponds to an error of the order of $6 \times 10^{-7}$ at the $1\sigma$ level on 
the model parameter $\fr = 5 \times 10^{-6}$.
We report also forecasted constraints on a model with a large value of $\fr$, 
namely $\fr = 5 \times 10^{-5}$ and a model closer to standard gravity with $\fr = 5 \times 10^{-7}$ 
and show that in the optimistic scenario, \
Euclid will be able to distinguish these models from \lcdm\ at more than 3$\sigma$.} 
{We find that with a  good control of systematic effects and  modelling of the matter power spectrum in the mildly and deeply non-linear regime,  \Euclid will be a powerful probe for $f(R)$ models. It will constrain the scale-dependence of the perturbations as well as a substantial part of the non-linear regime of structure formation, discerning these models from \lcdm\ .}

\keywords{
Cosmology: theory; large-scale structure of Universe; cosmological parameters; dark energy. Gravitational lensing: weak}

\maketitle

\section{Introduction}\label{sec:intro}

The origin of the accelerated expansion of the Universe is still challenging our understanding of late-time cosmology. A cosmological constant, $\Lambda$, remains in agreement with current data but its value, when considered as vacuum energy, does not correspond to theoretical predictions and is rather considered as a phenomenological parameter that fits the data. 
An appealing proposal for an alternative modelling  is that of modifying gravitational interactions felt by particles, 
either in a universal (same interaction for all particles) or non-universal way (acting differently on different particles). In this paper, we investigate one popular scenario which belongs to the first class, in which the theory of general relativity is modified by extending the Ricci scalar $R$ in the Hilbert-Einstein action with a general function of it, $R \rightarrow R+f(R)$. We forecast the ability that the forthcoming \Euclid satellite will have to constrain this scenario, via galaxy clustering (GC, photometric \GCph\ and spectroscopic \GCsp), weak lensing (WL), or their combination, either of the photometric probes alone (\XCph) or all together. In particular, we produce for the first time validated forecasts on the Hu-Sawicki $f(R)$ model \citep{Hu:2007nk}, whose background expansion mimics that of a cosmological constant model, while differing at the level of cosmological perturbations: the growth of structure is driven here by a modification of gravity (MG).

\Euclid\footnote{\url{http://www.euclid-ec.org/}} is a European Space Agency medium-class space mission due for launch in 2023. It will carry on-board a near-infrared spectrophotometric instrument \citep{NISP} and a visible imager \citep{VIS} that will allow it to perform both a spectroscopic and a photometric survey over $15\,000\,\deg^2$ of extra-Galactic sky \citep{Redbook}. The main aim of the mission is to measure the geometry of the Universe and the growth of structures up to redshift $z\sim 2$ and beyond.

\Euclid will include a photometric survey, measuring positions and shapes of over a billion galaxies, enabling the analysis of WL and \GCph. Given the relatively large redshift uncertainties that we expect from photometric measurements (compared to spectroscopic observations), these analyses will be performed via a tomographic approach, in which galaxies are binned into redshift slices that are considered as two-dimensional (projected) data sets. On the other hand, the spectroscopic survey will provide very precise radial measurements of the position of galaxies. Even if the number density will be lower -- compared to the photometric survey -- it will allow us to perform a galaxy clustering analysis in three dimensions, \GCsp.
The combination of photometric and spectroscopic surveys will enable a powerful test of the two independent gravitational potentials that are predicted to be different within $f(R)$ cosmologies.

We want to quantify the effect of combining the complementary information obtained from the two probes. It has been shown in \citet[][EC19 hereafter]{Blanchard:2019oqi} that combining \GCph\ and its cross-correlation (\XCph) with WL is able to improve the figure of merit by a factor of three, for dynamical dark energy models. Our goal here is to explore the impact of cross-correlation on the additional parameter $f_{R0} \equiv \mathrm{d}f/\mathrm{d}R\, (z=0)$ describing the standard model extensions within the $f(R)$ model, as will be defined in the next section. 

Several studies have tried to constrain the Hu-Sawicki model. Among them, \cite{Hu:2016zrh} used Planck15 cosmic microwave background (CMB) data, baryon acoustic oscillations (BAO), Supernovae Ia from JLA, WiggleZ and CFHTLenS data sets and obtained an upper bound of $|f_{R0}| < 6.3 \times 10^{-4}$ at the 95\% confidence level (C.L.). Similar constraints from cosmological data were obtained for instance in \citet{2017JCAP...01..005N}, \citet{2013PhRvD..87j3002O} and \citet{2018PhRvD..97b3525P}. A review of local and astrophysical 
constraints on $f_{R0}$ is given by \citet{Lombriser:2014dua}. 
Notably, a bound of $|f_{R0}| < 10^{-6}$ is obtained assuming that the Milky Way can be treated as an isolated system in the cosmological background with no environmental screening. Under similar assumptions, astrophysical tests are capable of constraining $|f_{R0}|$ at the level of $10^{-7}$ \citep{2019arXiv190803430B}. Furthermore \citet{Desmond2020} constrained $|f_{R0}|<1.4\times 10^{-8}$ using galaxy morphology.

After reviewing the $f(R)$ formalism in \cref{sec:fR}, we present in \cref{sec:thpred} the \Euclid primary probes, WL, $\GCph$ and \XCph, for the photometric part and $\GCsp$ for the spectroscopic part. We then present the survey specifications and analysis scheme in \cref{sec:fisher}. Finally, we present our results for the considered fiducial models in \cref{sec:results} and conclude in \cref{sec:conclusions}.

%----------------------------------------------------

\section{Hu-Sawicki $f(R)$ gravity}\label{sec:fR}
%----------------------------------------------------

A modification of Einstein's theory of general relativity (GR) can be obtained by promoting the linear dependence of the Hilbert-Einstein action, $S$, on the Ricci scalar $R$ to a non-linear function $R+f(R)$ \citep{1970MNRAS.150....1B},
\begin{align}
 S = \frac{c^4}{16\pi G_{\rm N}} \int{\de^4 x \sqrt{-g} \left[R+f(R)\right]} \,, \label{eq:EHaction}
\end{align}
where $g_{\mu\nu}$ is the metric tensor, $G_{\rm N}$ is Newton's gravitational constant and we have expressed explicitly the speed of light $c$, to allow for consistency with the choice of units in the observable quantities below.

The $f(R)$ family of cosmologies implies a universal coupling with all matter species, inducing an additional `fifth force'. Therefore, an important attribute that a viable late-time $f(R)$ modification must possess to leave a detectable signature in the cosmic structure formation while complying with stringent constraints on gravity in the Solar System, is that the functional form $f(R)$ gives rise to a screening mechanism, the so-called `chameleon mechanism' \citep{Khoury:2003aq}.

The fifth force has a range determined by the Compton wavelength $\lambda_C$ which has a very direct relation with the parameter $f_{R0}$. For cosmological densities one has $\lambda_C =32\sqrt{|f_{R0}|/10^{-4}}$ Mpc (\cite{Hu:2007nk} and \cite{Cabre:2012tq}). This relation is important because it is the screening scale that prevents us to have $f_{R0}=0$ as the fiducial model and thus being able to forecast what is the minimum $f_{R0}$ which is detectable.

\begin{figure}
 \centering
 \includegraphics[width=0.97\linewidth]{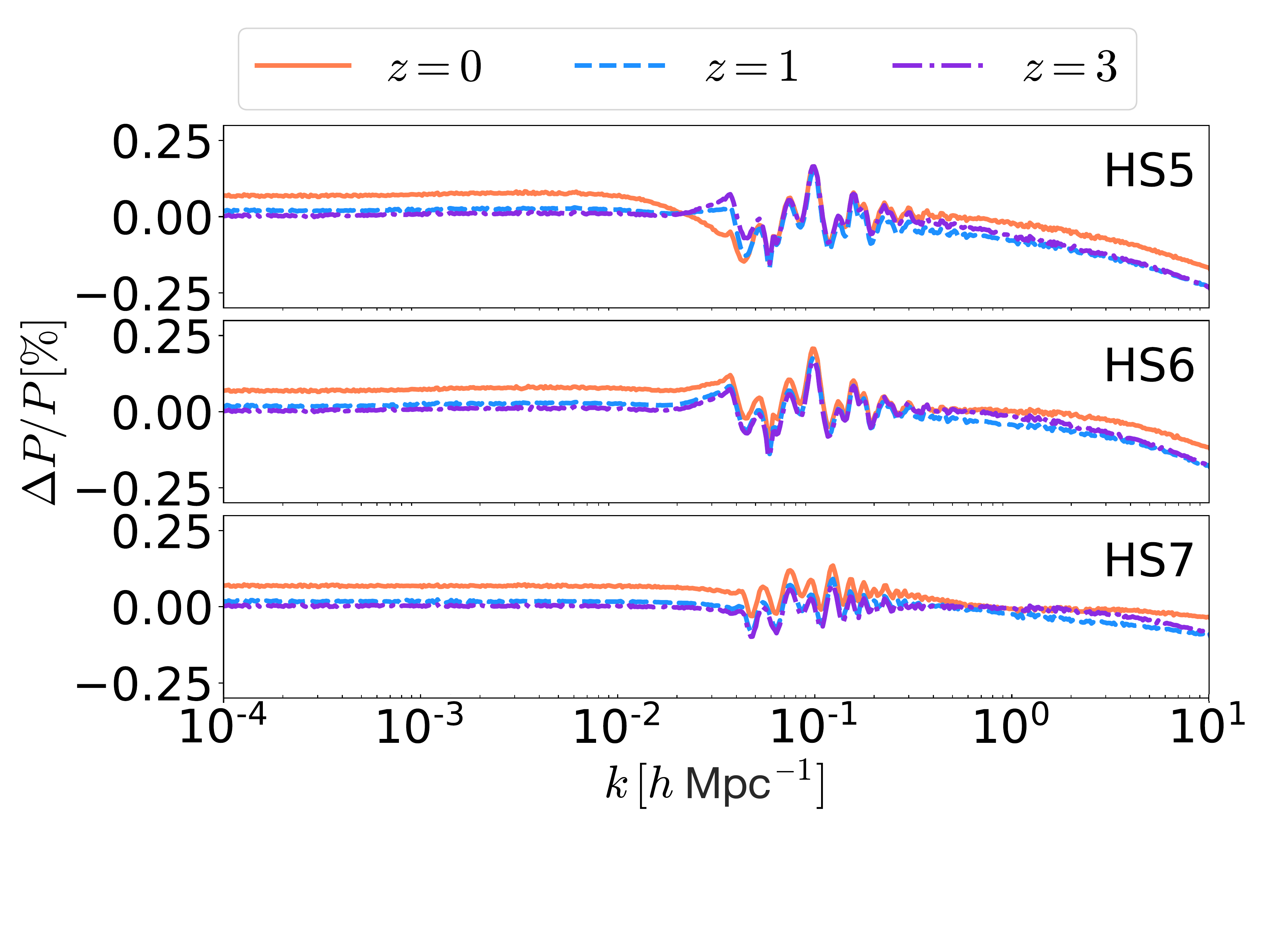}
 \caption{Relative difference between \texttt{EFTCAMB} and \texttt{MGCAMB} for the linear matter power spectrum ($\Delta P/P\equiv(P_{\texttt{EFTCAMB}}-P_\texttt{MGCAMB})/P_\texttt{MGCAMB}$) at three different redshifts, $z=0$ (solid orange line), $z=1$ (dashed blue line) and $z=3$ (dot-dashed purple line), for the three fiducial models with $|f_{R0}|= 5\times 10^{-5}$ (HS5), $|f_{R0}|= 5\times 10^{-6}$ (HS6) and $|f_{R0}|= 5\times 10^{-7}$ (HS7).}
 \label{fig:Comparison}
\end{figure}

As a specific example of this class of theories, we consider here the model proposed by \citet{Hu:2007nk}.
The functional form of $f(R)$, adopting $n=1$ for simplicity and the limit $|f_R|\ll1$, is given by
\begin{equation}
f(R) = - 6 \OmegaDE \frac{H_0^2}{c^2} + |f_{R0}| \frac{\bar{R}_0^2}{R}\,,\label{eq:fR}
\end{equation}
where $f_{R0} < 0$,
$\bar R_0$ denoting the Ricci scalar in the cosmological background today, $H_0$ the Hubble constant, and $\OmegaDE$ is the current fractional energy density attributed to a cosmological constant. The only additional free parameter of the model over \lcdm\ is therefore $f_{R0}$.

For the $|f_{R0}|\ll1$ values of interest here, the background expansion history approximates that of \lcdm\ and
\begin{equation}
\bar{R}_0 = 3 \Omegam \frac{H_0^2}{c^2} \left(1+ 4 \frac{\OmegaDE}{\Omegam} \right)\,,\label{eq:R}
\end{equation}
with matter energy density parameter $\Omegam=1-\OmegaDE$. 
It characterises the magnitude of the deviation from \lcdm\,, with smaller $|f_{R0}|$ values corresponding to weaker departures from GR. \lcdm\ is recovered in the limit of $f_{R0}\rightarrow0$.

At the perturbation level, deviations from GR can be encoded in phenomenological functions of the metric \citep{Zhang:2007nk, Amendola:2007rr, 2016A&A...594A..14P}.
Using the Bardeen formalism \citep{1995ApJ...455....7M}, we can define the conformal metric of the infinitesimal line element, $\de s$, in an expanding Universe as
\begin{equation}
 \de s^2 = a^2(\tau) \left[ - (1+2\Psi)\,c^2\,\de \tau^2 + (1-2\Phi)\,\de x^i\,\de x_i \right]\,,
\end{equation}
where $a(\tau)$ is the scale factor in conformal time, $\tau$, $\de x^i$ is the three dimensional infinitesimal spatial element, and $\Psi$ and $\Phi$ are the two scalar potentials. Then the phenomenological functions describe the modifications to the Poisson equations, namely% 

\begin{align}
-k^2\Psi & = \frac{4\pi\,G_{\rm N}}{c^2} \,a^2\mu\left[\bar\rho\Delta+3\left(\bar{\rho}+\frac{\bar{p}}{c^2}\right)\sigma\right]\,, \label{eq:mu}\\ 
%  \label{eq:mu}
k^2\left(\Phi-\eta\Psi\right) & = \frac{12\pi\,G_{\rm N}}{c^2}\,a^2\mu\left(\bar{\rho}+\frac{\bar{p}}{c^2}\right)\sigma\,, \\ 
%  \label{eq:eta}
-k^2\left(\Phi+\Psi\right) & = \frac{8\pi\,G_{\rm N}}{c^2}\,a^2\left\{\Sigma\left[\bar{\rho}\Delta+3\left(\bar{\rho}+\frac{\bar{p}}{c^2}\right)\sigma\right]-\right.\nonumber \\
& \left. \qquad\qquad\quad \frac{3}{2}\mu\left(\bar{\rho}+\frac{\bar{p}}{c^2}\right)\sigma\right\}\,, \label{eq:sigma}
\end{align}
where the background quantities $\bar{\rho}$ and $\bar{p}$ are respectively the density and pressure of the matter species and are only a function of time, whereas perturbations are functions of time and scale, $\sigma$ is the matter anisotropic stress,
$\bar\rho\Delta=\bar\rho\delta+3(aH/k)(\bar{\rho}+\bar{p}/c^2)v$ is the comoving density perturbation, with $\delta=\rho/\bar{\rho}-1$ the density contrast, and $v$ is the velocity potential. 
The phenomenological functions $\mu(a,k),\,\eta(a,k),\,\Sigma(a,k)$ are identically equal to 1 in the GR limit. 
Notice that only two of them are independent from each other, the third one being a combination of the other two; in the limit of negligible anisotropic stress from matter, the relation reduces to 
\begin{equation}
 \Sigma(a,k) = \frac{\mu(a,k)}{2}\left[1+\eta(a,k)\right]\,.
\end{equation}

These phenomenological functions 
can be determined analytically considering the quasi-static limit (i.e.\ scales sufficiently small to be well within the horizon and the sound-horizon of the scalar field). In the case of $f(R)$ gravity, the expressions 
reflect the presence of an additional fifth force with a characteristic mass scale \begin{equation}\label{eq:mass_fR}
m_{f_R}^2\sim \frac{1+f_R}{3f_{RR}}\sim \frac{1}{3f_{RR}}\,.
\end{equation}

For negligible matter anisotropic stress, one finds \citep{Pogosian:2007sw}
\begin{align}
 \mu(a,k) & = \frac{1}{1+f_R(a)}\frac {1+4k^2a^{-2}m_{f_R}^{-2}(a) }{1+3k^2a^{-2}m_{f_R}^{-2}(a)}\,,\label{eq:mu_fR}
\end{align}
and for the Hu-Sawicki model under consideration, $m_{f_R}$ is given by \citep{Brax:2013fna}
\begin{equation}
 m_{f_R}(a) = \frac{H_0}{c\sqrt{2|f_{R0}|}}\frac{\left(4\OmegaDE+\Omegam a^{-3}\right)^{3/2}}{4\OmegaDE+\Omegam}\,.
\end{equation}
Since $f(R)$ models have a conformal coupling, light deflection  is weakly affected as follows
\begin{equation}\label{eq:SigmaHS}
 \Sigma(a)=\frac{1}{1+f_{R}(a)}\,,
\end{equation}
and weak lensing is affected in the same way as matter growth, but with a different weight in time and scale.

This approach is at the basis of the Einstein-Boltzmann solver \texttt{MGCAMB} \citep{Zhao:2008bn,Hojjati:2011ix,Zucca:2019xhg}, or \texttt{MGCLASS} \cite{Baker:2015bva, Sakr:2021ylx}, each a modification of the standard Einstein-Boltzmann solver \texttt{CAMB} or \texttt{CLASS}, respectively.
Note that here we assume the pre-factor $1/(1+f_R)$ in \cref{eq:mu_fR} to be unity \citep{Hojjati:2015ojt}. 
This approximation is valid for viable values of $f_{R0}$. Given our choice of the fiducial values, $|f_{R0}|\ll1$, the deviation of $\Sigma$ from unity is also negligible.

Alternatively, the phenomenological functions, $\mu$, $\eta$, and $\Sigma$, can be determined numerically, after solving for the full dynamics of linear perturbations via \texttt{EFTCAMB} \citep{Hu:2013twa,Raveri:2014cka}, which implements the effective field theory formalism for dark energy into the standard \texttt{CAMB} code \citep{Lewis:1999bs}; see \citet{Hu:2016zrh} for an application to Hu-Sawicki $f(R)$ gravity. This code has been validated as part of an extended code comparison effort \citep{Bellini:2017avd}.

For the model under consideration, we have compared predictions of the angular power spectrum up to $\ell=5000$ of the CMB and the matter power spectrum up to $k=10\,h\,\mathrm{Mpc}^{-1}$ both from \texttt{MGCAMB} (quasi-static) and \texttt{EFTCAMB} (full evolution). Both codes lead to a sub-percent agreement, well within the desired level of accuracy. For the range of values of $|f_{R0}|$ considered in this work the agreement of the angular power spectra is never worse than $0.25\%$ for the temperature-temperature power spectrum (for $\ell<10^3$ it is below 0.1\%) and 0.1\% for the lensing power spectrum, for the matter power spectrum  the two codes agree extremely well up to $k = 0.02 \,h\,\mathrm{Mpc}^{-1}$ ($<0.1\%$) and for larger $k$ the relative difference is always below $0.25\%$. We show in Fig. \ref{fig:Comparison} for the matter power spectrum the relative difference between \texttt{EFTCAMB} and \texttt{MGCAMB} at three different redshifts. The agreement of the two codes has been tested against the  choice of GR transition time, i.e.\ the time at which a MG model starts to deviate significantly from its GR limit. We find that the level of agreement is not affected by this parameter (once this is the same in both codes). For the present analysis we set the GR transition time at $a=10^{-3}$. 
This choice is justified by considering that in the Hu-Sawicki model the growth function is scale dependent and large-$k$ modes show a significant deviation from GR already at high redshift. Moreover in this case an early GR transition time guarantees a smooth transition between the MG and GR regimes. 
We have also verified that this is the case using \texttt{MGCLASS}. Given the agreement of the codes, we conclude that the quasi-static approximation for the Hu-Sawicki $f(R)$ model is a valid assumption, and we proceed with the forecasts using the inputs from \texttt{MGCAMB}. 

\begin{figure}
 \centering
 \includegraphics[width=0.95\linewidth]{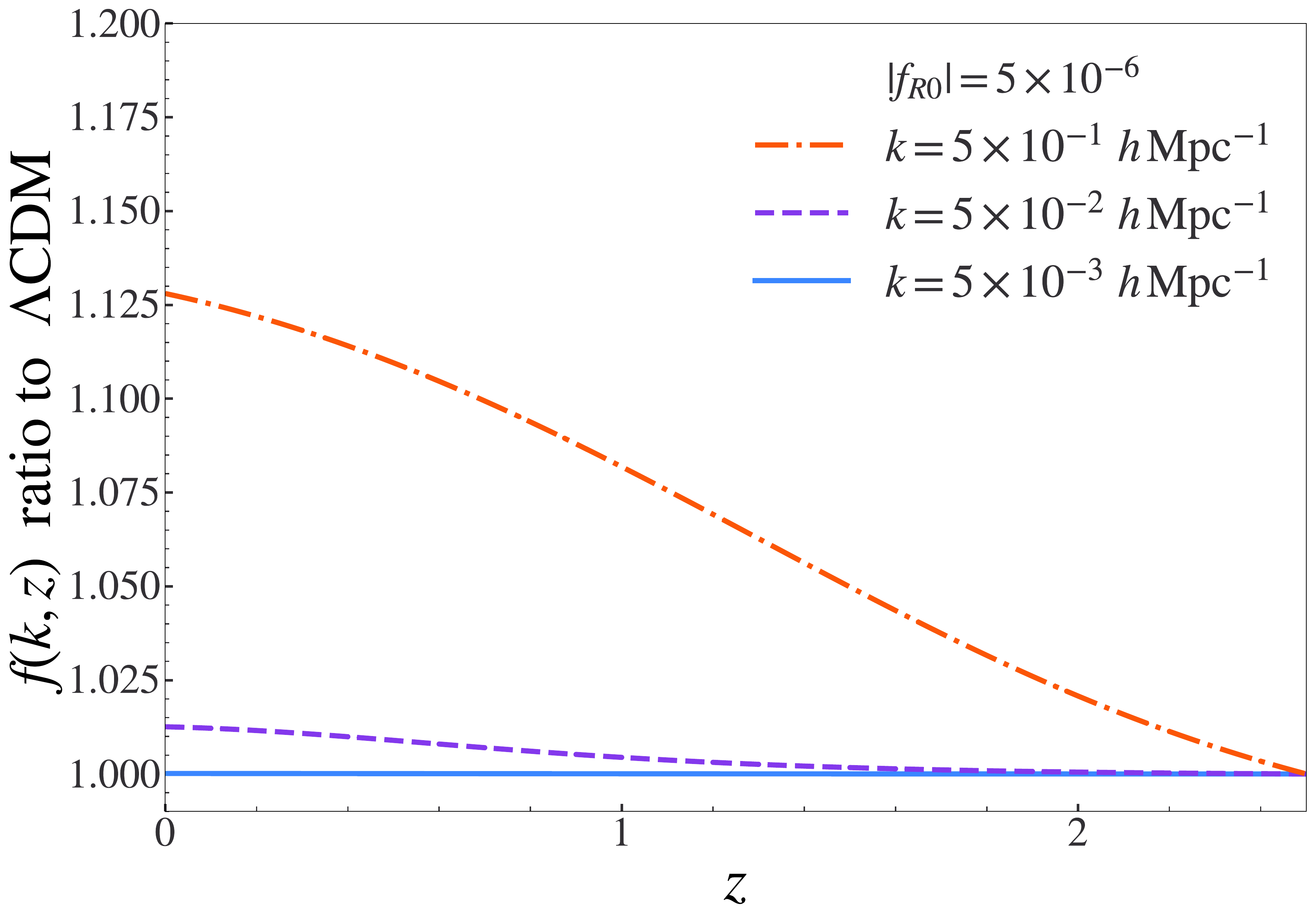}
 \caption{Ratio of the scale-dependent matter growth rate $f(k,z)$ in $f(R)$ gravity (for the fiducial value $\fr = 5 \times 10^{-6}$) 
 with respect to \lcdm, for three different wavenumbers $5 \times 10^{-3}$ (blue solid line), $5 \times 10^{-2}$ (dashed purple line) and $5 \times 10^{-1}\,h\,\mathrm{Mpc}^{-1}$ (dot-dashed orange line), as a function of redshift from $z=0$ to $z=2.5$.
 The smaller the spatial scales and the lower the redshifts the larger is the enhancement of the growth rate compared to \lcdm.}
 \label{fig:fr-scale-dep}
\end{figure}

Finally, the scale-dependent $\mu$ function in Eq. \eqref{eq:mu_fR}  introduces a scale-dependent growth of structures as shown by \citet{Zhang:2005vt} and \citet{Song:2006ej}. 
In \cref{fig:fr-scale-dep}, we plot the growth rate of perturbations, $f(k,a)\equiv \mathrm{d}\ln \delta/\mathrm{d}\ln a$, in $f(R)$ Hu-Sawicki gravity with respect to the one in \lcdm, for three different scales $k$, namely $5 \times 10^{-3}$ (blue solid line), $5 \times 10^{-2}$ (dashed purple line) and $5 \times 10^{-1}$ $h\,  \mathrm{Mpc}^{-1}$ (dot-dashed orange line), as a function of redshift from $z=0$ to $2.5$.
This shows that the growth of perturbations at large scales is very similar to the one in standard GR, at the redshifts of interests for large-scale structure formation. However, at smaller scales (larger $k$) the growth of perturbations is enhanced in $f(R)$ at low redshifts 
\cite[see][for a parametrization of the growth rate in generic $f(R)$ models, in terms of the growth index $\gamma$]{Buenobelloso:2011sja}. 
These very features also complicate the observational modelling, which moreover serves as a stress test of the forecasting pipeline for constraints on theories beyond \lcdm.

\section{Theoretical predictions for Euclid observables}\label{sec:thpred}

As it will be described in the next section, the forecasting methods and tools used in this paper are the same of \citetalias{Blanchard:2019oqi}. However, we must notice here that the change in the theory of gravity introduced through the Hu-Sawicki model implies significant modifications of the recipes used to compute theoretical predictions for \Euclid observables. We discuss in this Section how moving away from the standard GR assumption impacts the predictions for the angular power spectra $C(\ell)$ that will be compared with the photometric survey data, and on the power spectra $P_{\rm obs}$ compared with data of the spectroscopic survey.

\subsection{Photometric survey}\label{sec:photo}

For the \Euclid photometric survey, the observables that need to be computed and compared with the data are the angular power spectra for WL, $\GCph$ and their cross correlation, \XCph.

In \citetalias{Blanchard:2019oqi} these were calculated using the Limber approximation plus the flat-sky approximation with pre-factor set to unity in a flat \lcdm\ Universe, as 
\begin{equation}
 C^{XY}_{ij}(\ell) = \frac{c}{H_0}\int_{z_{\rm min}}^{z_{\rm max}}\de z\,{\frac{W_i^X(z)W_j^Y(z)}{E(z)r^2(z)}P_{\delta\delta}(k_\ell,z)} ,\label{eq:ISTrecipe}
\end{equation}
with $k_\ell=(\ell+1/2)/r(z)$, $r(z)$ the comoving distance to redshift $z=1/a-1$, $P_{\delta\delta}(k_\ell,z)$ the non-linear power spectrum of matter density fluctuations, $\delta$, at wave number $k_{\ell}$ and redshift $z$, in the redshift range of the integral from $z_{\rm min}=0.001$ to $z_{\rm max}=4$. 
The dimensionless Hubble function is defined as $E(z)=H(z)/H_0$ and in all subsequent equations $H_0$ is expressed in units of ${\rm km}\,{\rm s}^{-1}\,{\rm Mpc}^{-1}$.

For each tomographic redshift bin $i$, the window functions $W^X_i(z)$ with $X=\{{\rm L,G}\}$ (corresponding to WL and \GCph, respectively) need to be computed differently with respect to what was done in \citetalias{Blanchard:2019oqi} as, when abandoning the assumption of a GR gravity theory, one has to account for changes in the evolution of both the homogeneous background and of cosmological perturbations. However, in the case of the $f(R)$ model considered in this work, the background is $\lcdm$ up to a high precision.
In general, one also has to account for both the modified evolution of the Bardeen potentials, $\Phi$ and $\Psi$, and for the fact that in MG the GR relation $\Phi=\Psi$ is not necessarily satisfied. Using the modified Poisson equation for $\Phi+\Psi$ of \cref{eq:sigma}, this combination can be related to $P_{\delta\delta}$ as
\begin{equation}\label{eq:weylgen}
 P_{\Phi+\Psi}(k,z) = \left[-3\Omegam \left(\frac{H_0}{c}\right)^2(1+z)\Sigma(k,z)\right]^2P_{\delta\delta}(k,z)\,,
\end{equation}
where we assume a standard background evolution of the matter component, i.e.\ $\rho_{\rm m}(z)=\rho_{{\rm m},0}(1+z)^3$, and $P_{\delta\delta}$ is computed accounting for the MG effects introduced through \cref{eq:mu}.

We can, therefore, use the recipe of \cref{eq:ISTrecipe} accounting for the effects of these modifications of gravity, not general modifications, e.g. if matter coupling shifts $\Omega_{\rm m}(a)$. We can calculate $H$, $r$ and $P_{\delta\delta}$, provided by dedicated Boltzmann solvers, but with the new window functions \citep{SpurioMancini:2019rxy}
\begin{align}
 W_i^{\rm G}(k,z) =&\; \frac{H_0}{c}b_i(k,z)\frac{n_i(z)}{\bar{n}_i}E(z)\,, \label{eq:wg_mg}\\  
 W_i^{\rm L}(k,z) =&\; \frac{3}{2}\Omegam \left(\frac{H_0}{c}\right)^2(1+z)r(z)\Sigma(k,z) \,\times \nonumber\\
& \int_z^{z_{\rm max}}{\de z'\frac{n_i(z)}{\bar{n}_i}\frac{r(z'-z)}{r(z')}} + W^{\rm IA}_i(k,z)\, , \label{eq:wl_mg}
\end{align}
where $n_i(z)/\bar{n}_i$ and $b_i(k,z)$ are, respectively, the normalised galaxy distribution and the galaxy bias in the $i$-th redshift bin, and $W^{\rm IA}_i(k,z)$ encodes the contribution of intrinsic alignments (IA) to the WL power spectrum. We follow\,\citetalias{Blanchard:2019oqi} in assuming an effective scale-independent galaxy bias. The main reason for this choice is to be able to compare to the standard analysis with the modified gravity model as the only variable. Accounting for a scale-dependent galaxy bias would introduce further degrees of freedom that could confuse the comparison between the different cosmological models. A detailed analysis of both the concordance model and modified gravity theories accounting for scale-dependent galaxy bias is beyond the scope of this work\,\citep[see e.g.][for an analysis on the concordance model with a local, non-linear galaxy bias model]{Tutusaus2020}.

The IA contribution is computed following the eNLA model from \citetalias{Blanchard:2019oqi}, in which
\begin{equation}\label{eq:IA}
 W^{\rm IA}_i(k,z)=-\frac{\mathcal{A}_{\rm IA}\mathcal{C}_{\rm IA}\Omega_{\rm m,0}\mathcal{F}_{\rm IA}(z)}{\delta(k,z)/\delta(k,z=0)}\frac{n_i(z)}{\bar{n}_i(z)}\frac{H_0}{c}E(z)\,,
\end{equation}
where 
\begin{equation}
 \mathcal{F}_{\rm IA}(z)=(1+z)^{\eta_{\rm IA}}\left[\frac{\langle L\rangle(z)}{L_\star(z)}\right]^{\beta_{\rm IA}}\,,
\end{equation}
with $\langle L\rangle(z)$ and $L_\star(z)$ redshift-dependent mean and the characteristic luminosity of source galaxies as computed from the luminosity function, $\mathcal{A}_{\rm IA}$, $\beta_{\rm IA}$ and $\eta_{\rm IA}$ are the nuisance parameters of the model, and $\mathcal{C}_{\rm IA}$ is a constant accounting for dimensional units.

Changes in the theory of gravity impact the IA contribution introducing a scale dependence through the modified perturbations growth. This is explicitly taken into account in \cref{eq:IA} through the matter perturbation $\delta(k,z)$, which is considered to be scale dependent in this case. This allows us to consider also the scale dependence introduced by massive neutrinos, which was assumed to be negligible in \citetalias{Blanchard:2019oqi}.

Notice that while the window function for WL includes the MG function $\Sigma$ in order to properly account for the modifications to $\Phi+\Psi$, the $\GCph$ one does not have any explicit MG contribution, as the modifications on the clustering of matter are accounted for in the new $P_{\delta\delta}(k_\ell,z)$.

We can, therefore, apply this recipe to the Hu-Sawicki $f(R)$ model. Given our choice of the fiducial $f_{R0}$, discussed in \cref{sec:fisher}, the background modifications with respect to \lcdm\ are negligible, while this model affects the evolution of perturbations, and therefore $P_{\delta\delta}$, through \cref{eq:mu}, with the $\mu$ function given by \cref{eq:mu_fR}.

The function $\Sigma$ needs to relate the $\Phi+\Psi$ and matter power spectra, as in \cref{eq:weylgen}, and its is given by \cref{eq:SigmaHS}.
As previously discussed for our fiducial choice, $|f_{R0}|\ll1$, the deviations of $\Sigma$ from unity are negligible and therefore the geometrical part of lensing kernel entering \cref{eq:wl_mg} reduces to the standard one.

\subsection{Spectroscopic survey}\label{sec:spect}
In order to exploit data from the \Euclid spectroscopic survey, we need to compute the theoretical prediction for the observed galaxy power spectrum in the extended model considered here.

The full non-linear model for the observed galaxy power spectrum is given by 
\begin{multline}
P_\text{obs}(k_\text{ref},\mu_{\theta,\text{ref}};z) = 
\frac{1}{q_\perp^2(z) q_\parallel(z)} %AP
\left\{\frac{\left[b\sigma_8(k,z)+f\sigma_8(k,z)\mu_{\theta}^2\right]^2}{1+k^2\mu_{\theta}^2\sigma_{\rm p}^2(z)}\right\} 
\\
\times \frac{P_\text{dw}(k,\mu_{\theta};z)}{\sigma_8^2(z)}  %non-linear damping
F_z(k,\mu_{\theta};z) %z-error
+ P_\text{s}(z) \, , % shot noise
\label{eq:GC:pk-ext}
\end{multline}
where the $P_{\rm dw}(k,\mu;\,z)$ is the de-wiggled power spectrum which models the smearing of the BAO features due to the displacement field of wavelengths smaller than the BAO scale,
\begin{equation}
P_\text{dw}(k,\mu;z) = P_{\delta\delta}(k;z)\,\text{e}^{-g_\mu k^2} + P_\text{nw}(k;z)\left(1-\text{e}^{-g_\mu k^2}\right) \,,
\label{eq:GC:pk_dw}
\end{equation}
where the $P_{\rm nw}(k;z) $ is a ‘no-wiggle’ power spectrum with the same broad band shape as $P_{\delta\delta}(k;z) $ but without BAO features (see below for details on how we compute it).

In \cref{eq:GC:pk-ext}, $k$ is the modulus of the wave vector $\bm k$ and $\mu_{\theta}$ is the cosine of the angle $\theta$ between this vector and the line-of-sight direction $\hat{\bm r}$. These quantities on the right hand side are functions of their counterparts at a reference cosmology, i.e. $k \equiv k(k_{\rm ref})$, $\mu_{\theta} \equiv \mu_{\theta,\text{ref}}$, which are transformed due to the Alcock-Paczynski effect, see \citetalias{Blanchard:2019oqi} and \cite{Euclid:2023pxu} for the explicit formula. This transform, which also scales the overall $P_\text{obs}$ is parameterised in terms of the angular diameter distance $D_{\rm A}(z)$ and the Hubble parameter $H(z)$ as 
\begin{align}
q_{\perp}(z) &= \frac{D_{\rm A}(z)}{D_{\rm A,\, ref}(z)},\\
q_{\parallel}(z) &= \frac{H_\text{ref}(z)}{H(z)}\,.
\end{align}

The term in the curly brackets in \cref{eq:GC:pk-ext} is the contribution of redshift space distortions (RSD) corrected for the non-linear Finger-of-God (FoG) effect, where we defined $b\sigma_8(k,z)$ as the product of the effective scale-dependent bias of galaxy samples and the r.m.s.\ matter density fluctuation $\sigma_8(z)$; similarly, $f\sigma_8(k,z)$ is the product of the scale-dependent growth rate and $\sigma_8(z)$. As in the photometric survey, we follow\,\citetalias{Blanchard:2019oqi} in using an effective scale-independent galaxy bias. An analysis of modified gravity with scale-dependent galaxy bias models is left for future work.

The observed galaxy power spectrum is modulated by the redshift uncertainties which is manifested as a smearing of the galaxy density field along the line-of-sight, hence the factor $F_z$ in \cref{eq:GC:pk-ext} reads 
\begin{equation}
    F_z(k, \mu_{\theta};z) = \text{e}^{-k^2\mu_{\theta}^2\sigma_{r}^2(z)}\,,
\end{equation}
being $\sigma_{r}^2(z) = c(1+z)\sigma_{0,z}/H(z)$ and $\sigma_{0,z}$ is the error on the measured redshifts. 

Finally, the $P_{\rm s}(z)$ is a scale-independent shot noise term, which enters as a nuisance parameter \citepalias[see][]{Blanchard:2019oqi}.

The change in the gravity model affects the way to compute the theoretical predictions, as these expressions need to account for the possibility for the growth rate $f(z)$ to also depend on the wave number $k$. In general, this is the case for any MG model, and also when perturbation in the dark energy sector are considered. For the model considered in this paper, the only terms affected are those directly related to the growth rate, in brief: $f(k,z)$ itself, and the two phenomenological parameters related to the velocity dispersion, $\sigma_{\rm v}$, and the pairwise velocity dispersion, $\sigma_{\rm p}$. These are
\begin{align}
\sigma^2_{\rm v}(z, \mu_{\theta}) &= \frac{1}{6\pi^2}\int\de k\, P_{\delta\delta}(k,z)\left\{1 - \mu_{\theta}^2 + \mu_{\theta}^2\left[1+f(k,z)\right]^2\right\},\label{eq:sigmav}\\
\sigma_{\rm p}^2(z) &= \frac{1}{6\pi^2}\int\de k\, P_{\delta\delta}(k,z)f^2(k,z)\,.\label{eq:sigmap}
\end{align}

The phenomenological parameters $\sigma_{\rm v}(z,\mu_{\theta})$ and $\sigma_{\rm p}(z)$ account for the damping of the BAO features and the FoG effect, respectively. The smearing of the BAO peak is due to the bulk motion of scales smaller than the BAO scale. For the power spectrum, this can be modeled in the Zeldovich approximation by a multiplicative damping term of the form $\exp\left[-k^i k^j \langle d^i (z) d^j(z) \rangle\right]$,
where $\langle d^i (z) d^j(z) \rangle$ is the correlation function of the displacement field $d^i$ evaluated at zero distance (see for instance the Appendix~C of \citealt{Peloso:2015jua} for more details). Finally, we would like to clarify that in \cref{eq:GC:pk_dw} we used the function $g_{\mu}$ to express the damping of the BAO features in the matter power spectrum to keep the recipe closer to the \citetalias{Blanchard:2019oqi}; in this work, $g_\mu = \sigma_{\rm v}(z,\,\mu_{\theta})$.

In \lcdm\ model the growth rate is scale-independent and both \cref{eq:sigmap,eq:sigmav} are the same \citepalias[see][]{Blanchard:2019oqi}. Notice that even if these parameters are assumed to be the same, they come from two different physical effects, namely large-scale bulk flow for the former and virial motion for the latter.
Finally, due to the scale dependence of $\sigma_{\rm p}$ and $\sigma_{\rm v}$, we evaluated both parameters at each redshift bin but we kept them fixed in the Fisher matrix analysis. This method corresponds to the optimistic settings in \citetalias{Blanchard:2019oqi}. 
We would like to highlight that in this work we take directly the derivatives of the observed galaxy power spectrum with respect to the final parameters, contrary to \citetalias{Blanchard:2019oqi} where first we performed the Fisher matrix analysis for the redshift dependent parameters $H(z)$, $D_{\rm A}(z)$ and $f\sigma_8(z)$ and then projected to the final cosmological parameters of interest.

The no-wiggle matter power spectrum $P_{\rm nw}(k;z)$ entering \cref{eq:GC:pk_dw} has been obtained using a Savitzky-Golay filter to the matter power spectrum $P_{\delta\delta}(k;z)$. The Savitzky-Golay filter is usually applied to noisy data in order to smooth their behavior. This convolution method consists in fitting successive sub-sets of adjacent data points with a low-order polynomial. If the data are equally spaced (as it is in our case, equally spaced in $\log_{10}k$), then an analytic solution to the least-squares can be found as series of coefficients that can be applied to all the sub-sets. In practice, using the Savitzky-Golay filter we recover exactly the same shape and amplitude of the matter power spectrum without the BAO wiggles. While in \citetalias{Blanchard:2019oqi} we used the Eisenstein-Hu fitting formula \citep{Eisenstein:1997ik} for the no-wiggle power spectrum, this is a fitting formula that only applies approximately to $\Lambda\textrm{CDM}$ models and therefore cannot be straightforwardly applied in our case. We find that the Savitzky-Golay (SG) method is more accurate for models where the growth of matter density field depends on the scale $k$. The aforementioned smoothing filter has also been used in previous works \citep{Boyle:2017lzt} to reconstruct the neutrinos masses from galaxy redshift survey.
In \cref{fig:SGfilter-A1} we plot our reconstruction of the wiggles computed with the Eisenstein-Hu formula compared to the SG method used in this work, where we can see that it performs very well also in the case of $f(R)$.

\subsection{Non-linear modelling}\label{sec:non-linear}
While for the galaxy power spectrum on mildly non-linear scales we use a modified version of the model in \citetalias{Blanchard:2019oqi}, we do not have, in general, an analytical solution for the deeply non-linear power spectrum in a $f(R)$ cosmology. In this work, we will, therefore, use a fitting formula designed in \citet{Winther:2019mus} that captures the enhancement in the power spectrum compared to a \lcdm\ non-linear power spectrum, as a function of the parameter $f_{R0}$. This fitting function has been calibrated using the \texttt{DUSTGRAIN} \citep{Giocoli:2018gqh} and the ELEPHANT \citep{Cautun:2017tkc} $N$-body simulations \citep[see][for a comparison of different $N$-body codes for $f(R)$ cosmologies]{Winther:2015wla}.

\begin{figure}
 \centering
 \includegraphics[width=0.95\linewidth]{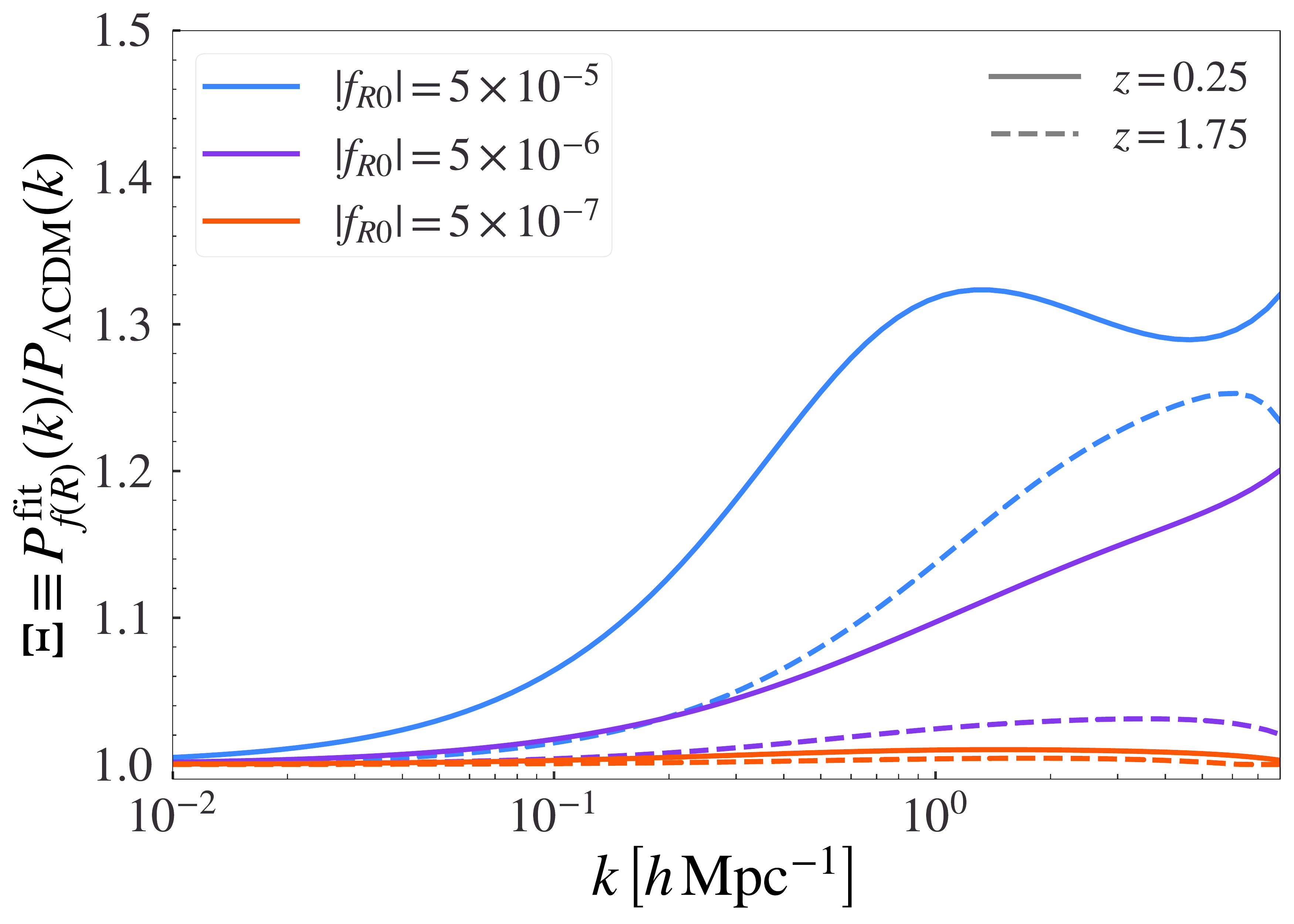}
 \caption{Ratio of the non-linear power spectrum in $f(R)$ gravity to $\lcdm$ from the fitting function in \cref{eq:fRfit} for three different values of the $\fr$ parameter, namely $5 \times 10^{-5}$ (blue line), $5 \times 10^{-6}$ (purple line) and $5 \times 10^{-7}$ (orange line), as a function of scale $k$, evaluated at two different redshifts $z=0.25$ (solid lines) and $z=1.75$ (dashed lines). These redshifts correspond approximately to the means 
 of the first and last lensing tomographic bins of our survey, respectively (see \cref{sec:fisher}).
 The fitting formula designed in \citet{Winther:2019mus} is not defined outside $10^{-4} < \fr < 10^{-7} $.} 
 \label{fig:fr-fitting-formula}
\end{figure}

The fitting function we use is given by
\begin{align}\label{eq:fRfit}
\frac{P^{\rm fit}_{f(R)}(k,z)}{P_{\lcdm}(k,z)} &= \Xi(k,z)\nonumber\\
&\equiv 1 + X_1\frac{1+ X_2\,k}{1+ X_3\,k}  \arctan(X_4\,k)^{X_5 + X_6\,k}\,,
\end{align}
where the $\{X_i\}$ are themselves functions of $f_{R0}$ and redshift, $X\equiv X(z;f_{R0})$.
The redshift dependence is a polynomial relation to the scale factor $a=1/(1+z)$ given by
\begin{align}
X_i(z;f_{R0}) = X_{i0}(y) + X_{i1}(y)(a-1) + X_{i2}(y)(a-1)^2 \,,
\end{align}
with each $X_{ij}$ coefficient in itself defined as a polynomial in  $y\equiv \log(f_{R0} / f_{R0}^{\rm fid})$, given by
\begin{align}
X_{ij}(y) = X_{ij0} + X_{ij1} y + X_{ij2}y^2  \,.
\end{align}
With three indices for $X_{ijk}$ this gives in total $6\times3\times3=54$ free parameters for the full scale, redshift and $f_{R0}$ dependence.
The response function $\Xi(k,z)$ is the ratio of the non-linear matter power spectrum in $f(R)$ theory to the non-linear spectrum calculated within the \lcdm\ model. The fitting formula \cref{eq:fRfit} found in \citet{Winther:2019mus} is a direct fit to this ratio, $\Xi(k,z)$. 
This fitting formula is not defined outside the range $10^{-4} < \fr < 10^{-7} $, therefore this will limit our smallest $\fr$ fiducial value across probes to be $5 \times 10^{-7}$, since we need to be far from the lower limit to be able to compute the numerical derivatives accurately.
In \cref{fig:fr-fitting-formula} we plot the function $\Xi(k,z)$ for each of the fiducial $\fr$ values chosen in this model, namely 
$5 \times 10^{-5}$ (blue line), $5 \times 10^{-6}$ (purple line) and $5 \times 10^{-7}$ (orange line), as a function of scale $k$. 
To compare the enhancement with respect to $\lcdm$ at our redshifts of interest, we evaluate it at $z=0.25$ (solid lines) and $z=1.75$ (dashed lines). These redshifts correspond approximately to the means of the first and last lensing tomographic bins of our survey, respectively (see \cref{sec:fisher}).
As can already be seen from this figure, the enhancement with respect to $\lcdm$ decreases rapidly with a smaller $\fr$ value, therefore we expect to have worst constraints for smaller values of $\fr$ when testing this model against probes that are sensitive to the deeply non-linear power spectrum.
Notice, however, that the Fisher matrix sensitivity is dominated by the response of this function to small changes in $\fr$ (i.e. its first derivatives), which can become large at small scales and therefore, we can get relatively tight constraints, even if the absolute enhancement over $\lcdm$ is of just a few percent evaluated at the fiducial. In \cref{sec:fisher} we will specify our choice of fiducial parameters for each model.

We implement $\Xi(k,z)$ into the Boltzmann codes \texttt{MGCAMB} and \texttt{EFTCAMB} and in order to obtain the non-linear $f(R)$ matter power spectrum, we then multiply $\Xi(k,z)$ by a \lcdm\ non-linear spectrum, $P^{\rm fit}_{f(R)}(k,z) = \Xi(k,z)\,P_{\lcdm}(k,z)$. For the \lcdm\ power spectrum $P_{\lcdm}(k,z)$, we use the Halofit `Takahashi' prescription \citep{Takahashi:2012em}, since this is the prescription most readily available in \texttt{MGCAMB}, and also the prescription used in \citet{Winther:2019mus} to test the fitting formula against $N$-body simulations. During the preparation of this work, an emulator for the deeply non-linear matter power spectrum has been developed by the \Euclid collaboration, the \texttt{EuclidEmulator} \citep[see][for details on its implementation]{Knabenhans:2018cng}, calibrated on the \Euclid Flagship simulation \citep{Potter2017}. Whilst it might be interesting to use it in the future, the current available version, \texttt{EuclidEmulator2} \citep[see][]{2020arXiv201011288E}, offers a \texttt{Python} wrapper to the Boltzmann code \texttt{CLASS} \citep{Lesgourgues:2011re, Blas:2011rf} rather than to \texttt{MGCAMB} and \texttt{EFTCAMB} we use for this paper; we are therefore not using the \texttt{EuclidEmulator} in the current analysis (see \cref{sec:fR}).
Also, during the preparation of this work, an emulator for the non-linear matter power spectrum in $\fr$ was developed by \citet{arnold2021forge} and goes under the acronym \texttt{FORGE}. It has been checked independently by the authors of this paper that this emulator agrees well with the fitting formula by \citet{Winther:2019mus} around the fiducial values of interest. However, this \texttt{FORGE} emulator has been calibrated with massless neutrino simulations and allows only variations of the $\Omegam$, $\sigma_8$ and $h$ cosmological parameters. Since in this work we want to have the flexibility to vary all other cosmological parameters and also the ability to connect to accurate modified gravity Boltzmann codes, we leave the application of this emulator for future work.
The fitting formula in \citet{Winther:2019mus} is also checked to be valid in the presence of non-zero neutrino masses, using data from \citet{Baldi_2014}. With the value $\sum m_\nu=0.06\,\mathrm{eV}$ chosen in this work, this fitting formula is accurate enough across our ranges of scales and redshifts of interest.

Note that in the present analysis, for the spectroscopic probe, we adopt the non-linear modelling described above in \cref{eq:GC:pk-ext} as it represents the minimal modification to what used in the \lcdm\ forecasts of \citetalias{Blanchard:2019oqi}. We choose to follow this more simplistic route, which allows  a clean comparison with \citetalias{Blanchard:2019oqi} with fixed theoretical systematics. However, it is worth noting that the state of the art, based e.g.\ on the Taruya-Nishimichi-Saito model (\citeyear{Taruya:2010mx}) or on perturbation theory prescriptions with bias expansion, has been formulated and recently implemented in the analysis of real data from the BOSS survey \citep{2015PhRvD..92d3522S, 2020JCAP...06..001C}.
Furthermore, eBOSS analyses \citep{2014MNRAS.443.1065B,2017MNRAS.466.2242B,2021MNRAS.501.5616D}, as well as forecasts for unbiased parameter estimation for Stage IV cosmological surveys, show that the choice of $k_{\rm max}$ is not universal \citep[see e.g.][]{2019OJAp....2E..13M}, and different $k_{\rm max}$ for the monopole, quadrupole, and hexadecapole are required (the $k_{\rm max}$ for the latter being considerably smaller). Therefore, we study several $k_{\rm max}$ choices, for the full shape of the power spectrum of spectroscopically-observed galaxies.

For the photometric observables, on the other hand, we are probing up to smaller scales. For this reason, including nuisance parameters about baryonic feedback on the matter power spectrum would be necessary for unbiased parameter estimation, and it would also possibly entail a degradation of the constraints from WL \citep[see e.g.][]{Schneider:2019xpf,2020JCAP...04..019S}. However, at the moment we do not have accurate \Euclid-like simulations including baryonic effects, especially in the case of MG cosmologies. Therefore, we ignore these effects in our analysis, leaving their inclusion for a future work.

We are aware that there are important degeneracies among the effect of $f(R)$ and neutrino masses, especially at non-linear scales \citep[see e.g.][]{Hu_2015, Baldi_2014}. While on the one hand massive neutrinos suppress the power spectrum at small scales, $f(R)$ will increase clustering at a similar range of scales, therefore partially cancelling the former effect \citep{Baldi_2014}. Ignoring these degeneracies might indeed artificially tighten our constraints on $\fr$ since we would be considering a much higher signal than the one actually present under a large sum of neutrino masses \citep{He_2013, Motohashi_2013, Harnois_D_raps_2015}. 
Breaking these degeneracies is possible, either by using statistics on the cosmic web \citep{shim2014breaking}, higher than second order statistics in weak lensing \citep{Peel_2018, Giocoli:2018gqh} or machine learning \citep{Peel_2019,Merten_2019}. These techniques are however beyond the scope of this work. Galaxy clusters and voids also offer a possibility to distinguish between these two possible scenarios, as investigated in \citet{Hagstotz_2019, Hagstotz_2019_2, ryu2020breaking, contarini2020cosmic}.
For our spectroscopic observable, $\GCsp$, there is also the possibility of breaking this degeneracy, by using information on redshift space distortions. This involves extending the perturbation theory modelling \citep{Wright_2019} or performing a detailed comparison with galaxy mocks from simulations \citep{Garc_a_Farieta_2019}. An exploration of these extensions on our non-linear modelling of RSD will be left for a future investigation.

\section{Survey specifications and analysis method}\label{sec:fisher}
In order to forecast constraints on this specific model, we follow the same approach of \citetalias{Blanchard:2019oqi}. We adopt the same Fisher matrix formalism, as well as the codes validated therein. Given that the theoretical model considered is crucially different, we update the forecast recipe and the corresponding codes as described in the previous section. The cosmological parameters here considered, and their fiducial values, for which we again follow \citetalias{Blanchard:2019oqi}, read
\begin{align}
    &\bm\Theta \phantom{_{\rm fid,1}}=\{\Omega_{\rm m,0},\, \Omega_{\rm b,0},\, h,\, n_{\rm s},\, \sigma_8,\, \logfr\}\,,\nonumber\\ 
    &\mathrm{HS5}: \bm\Theta_{\rm fid,HS5}=\{ 0.32,\, 0.05,\, 0.67,\, 0.96,\, 0.911,\, -4.301\}\,,\nonumber \,\\
    &\mathrm{HS6}:\bm\Theta_{\rm fid,HS6}=\{ 0.32,\, 0.05,\, 0.67,\, 0.96,\, 0.853,\, -5.301\}\,,\nonumber \,\\
    &\mathrm{HS7}:\bm\Theta_{\rm fid,HS7}=\{ 0.32,\, 0.05,\, 0.67,\, 0.96,\, 0.823,\, -6.301\}\,.
    \label{eq:fiducial-params}
\end{align}

The fiducial values of $\sigma_8$ shown above are obtained keeping the scalar amplitude of the primordial power spectrum $A_{\rm s}$ fixed for all three cosmologies. Given the impact of the $f(R)$ model under examination, this leads to $\sigma_8$ values that appear in tension with currently available results. However, it is important to stress that the comparison should be done with modified gravity analyses of present data, rather than with results obtained assuming $\Lambda$CDM. Indeed, the fiducial values of $\sigma_8$ we quote are compatible with such analyses, as degeneracies between $\sigma_8$ and modified gravity parameters increase the mean value of the former while also broadening the constraints \citep[see e.g.][]{DES:2022ccp}.

In the following text and in the figures we will refer to the model with the corresponding $|f_{R0}|=5\times10^{-5}$ fiducial as HS5, the model with $|f_{R0}|=5\times10^{-6}$ as HS6 and the model with a fiducial value of $|f_{R0}|=5\times10^{-7}$ as HS7. The baseline fiducial used for the rest of this work will be the HS6 model, since it corresponds to a value of $\fr$ still allowed by observations and with enough distinctive signatures with respect to $\lcdm$ as to be detected by future observations.
Note that our fiducial cosmology includes massive neutrinos with total mass of $\sum m_\nu=0.06\,\mathrm{eV}$, but we keep $\sum m_\nu$ fixed in the following Fisher matrix analysis. We also use the same initial amplitude of primordial perturbations for both fiducial models, namely $A_{\rm s} = 2.12605 \times 10^{-9}$. As discussed previously in \cref{sec:non-linear}, fixing neutrino masses ignores the degeneracies between their effect on the power spectrum and the increase of clustering at small scales, therefore our constraints might be tighter than in a scenario in which also $\sum m_\nu$ is varied.

Concerning the photometric probes, the galaxy distribution is binned into $10$ equi-populated redshift bins with an overall distribution following
\begin{equation}
    n(z)\propto\left(\frac{z}{z_0}\right)^2\,\text{exp}\left[-\left(\frac{z}{z_0}\right)^{3/2}\right]\,,
\end{equation}
with $z_0=0.9/\sqrt{2}$ and the normalisation set by the requirement that the surface density of galaxies is $\bar{n}_g=30\,\mathrm{arcmin}^{-2}$. The redshift distribution is then convolved with a sum of two Gaussian distributions to account for the photometric redshift uncertainties \citepalias[see][for details]{Blanchard:2019oqi}. The galaxy bias is assumed to be constant within each redshift bin, and its values $b_i$ are introduced as nuisance parameters in our analysis, with their fiducial values determined by $b_i = \sqrt{1+\bar{z}_i}$, where $\bar{z}_i$ is the mean redshift of each redshift bin. Even though deviations from GR introduce in principle a scale dependence also in the galaxy bias, we assume that this is negligible in our case.

Moreover, we follow \citetalias{Blanchard:2019oqi} in accounting for a Gaussian covariance between the different photometric probes:
\begin{multline}
    \text{Cov}\left[C_{ij}^{AB}(\ell),C_{kl}^{CD}(\ell')\right]=\frac{\delta_{\ell\ell'}^{\rm K}}{(2\ell+1)f_{\rm sky}\Delta \ell}\\
    \times\left\{\left[C_{ik}^{AC}(\ell)+N_{ik}^{AC}(\ell)\right]\left[C_{jl}^{BD}(\ell')+N_{jl}^{BD}(\ell')\right]\right.\\
    +\left.\left[C_{il}^{AD}(\ell)+N_{il}^{AD}(\ell)\right]\left[C_{jk}^{BC}(\ell')+N_{jk}^{BC}(\ell')\right]\right\}\,,
\end{multline}
where upper-case Latin indexes $A,\ldots=\{{\rm WL},\,\GCph\}$, lower-case Latin indexes $i,\ldots$ run over all tomographic bins, $\delta_{\ell\ell'}^{\rm K}$ is the Kronecker delta symbol, $f_{\rm sky}\simeq0.36$ represents the fraction of the sky observed by \Euclid, and $\Delta \ell$ denotes the width of the multipole bins, where we use $100$ equi-spaced bins in log-space. The noise terms, which for the observables considered here are in fact white noise, namely $N_{ij}^{AB}(\ell)\equiv N_{ij}^{AB}$, read
\begin{align}
    N_{ij}^{\rm LL}(\ell) &= \frac{\delta_{ij}^{\rm K}}{\bar{n}_i}\sigma_\epsilon^2\,,\\
    N_{ij}^{\rm GG}(\ell) &= \frac{\delta_{ij}^{\rm K}}{\bar{n}_i}\,,\\
    N_{ij}^{\rm GL}(\ell) &= 0\,,
\end{align}
where $\sigma_\epsilon^2=0.3^2$ is the variance of observed ellipticites.

For the spectroscopic probe, we evaluate the Fisher matrix $F_{\alpha\beta}(z_i)$ for the 
observed galaxy power spectrum according to the recipe outlined in \citetalias{Blanchard:2019oqi} (see Section~3.2). Here, $\alpha$ and $\beta$ run over the cosmological parameters of the set $\bm\Theta$, the index $i$ labels the redshift bin, each respectively centred in $z_i = \{1.0,\,1.2,\,1.4,\,1.65\}$, whose widths are $\Delta z = 0.2$ for the first three bins and $\Delta z = 0.3$ for the last bin. In this paper, in comparison to \citetalias{Blanchard:2019oqi}, we adopt the direct derivative approach, i.e.\ we vary the observed galaxy power spectrum with respect to the cosmological parameters of $\bm\Theta$ directly, plus two additional redshift-dependent parameters $\ln b\sigma_8(z_i)$ and $P_{\rm s}(z_i)$ that we marginalise over. We consider the numerical values for the galaxy bias, $b(z)$, and the expected number density of the observed H$\alpha$ emitters, $n(z)$,  reported in table 3 of \citetalias{Blanchard:2019oqi}.

For both probes, we consider two different scenarios: an optimistic and a pessimistic case. In the optimistic case, we consider $k_{\rm max}=0.30\,h\,\mathrm{Mpc}^{-1}$ for \GCsp, $\ell_{\rm max}=5000$ for WL, and $\ell_{\rm max}=3000$ for $\GCph$ and \XCph. Instead, in the pessimistic scenario, we consider $k_{\rm max}=0.25\,h\,\mathrm{Mpc}^{-1}$ for \GCsp, $\ell_{\rm max}=1500$ for WL, and $\ell_{\rm max}=750$ for $\GCph$ and \XCph. 
At the smallest photometric redshift bin, the galaxy number density distribution $n(z)$ peaks at around $z=0.25$, which means that under the Limber approximation and for our fiducial cosmology, the corresponding maximum values of $k$ evaluated in the power spectrum corresponding to the pessimistic and optimistic scenarios for \GCph\ are $k_{\rm max}=[0.7, 2.9]\,h\,\mathrm{Mpc}^{-1}$, respectively and for WL the maximum wavenumbers probed are $k_{\rm max}=[1.4, 4.8]\,h\,\mathrm{Mpc}^{-1}$ for pessimistic and optimistic, respectively. For smaller values of $z$, the values of $k$ at a given $\ell$ increase monotonically, but there the window functions in \cref{eq:wg_mg} and \cref{eq:wl_mg} suppress the power spectrum and we set it to zero after a fixed $k_{\rm max}=30\,h\,\mathrm{Mpc}^{-1}$.
In both scenarios, we fix the $\sigma_{\rm p}$ and $\sigma_{\rm v}$ nuisance parameters for $\GCsp$, which we calculate directly from \cref{eq:sigmap,eq:sigmav} for the fiducial value of the cosmological parameters. We also perform a second pessimistic forecast for \GCsp\ only, where we set our maximum wave number at $k_{\rm max} = 0.15\,h\,{\rm Mpc}^{-1}$ in order to have a more conservative estimate of the constraining power of the \GCsp\ probe. The reason for this is that the underlying matter power spectrum of \cref{eq:GC:pk-ext} that we are using in our observed galaxy power spectrum recipe is a linear one, as we detailed in \cref{sec:spect}. It is known that non-linear corrections start playing a role above scales of around $k = 0.1\,h\,{\rm Mpc}^{-1}$ for the redshifts under consideration \citep[see][]{Taruya:2010mx} and, therefore, the use of a linear power spectrum beyond these scales can bias our constraints. Hence, we aim to estimate what would happen if we used just quasi-linear scales in the analysis. As shown in \cref{sec:results}, the effect of these two different scale cuts on the constraining power on the $\logfr$ parameter is minimal for our \GCsp\ recipe. As a reference for the reader, we list the specific choices of scales and settings used for each observable in \cref{tab:specifications-ec-survey}.
\begin{table}
	\centering
	\caption{\Euclid survey specifications for WL, $\GCph$ and $\GCsp$. }
	\label{tab:specifications-ec-survey}
	\begin{tabularx}{\columnwidth}{Xll}
		\hline 
% 		& Parameter  & \Euclid \\
% 		\hline \hline
		Survey area & $A_{\rm survey}$  & $15\,000\,\deg^2$  \\
		\hline
		\hline
		\multicolumn{3}{c}{WL}\\
		\hline
		Number of photo-$z$ bins & $N_z$ & 10 \\
		Galaxy number density & $\bar n_{\rm gal}$  & $30\,\mathrm{arcmin}^{-2}$ \\
		Intrinsic ellipticity $\sigma$ & $\sigma_\epsilon$  & 0.30 \\
		Minimum multipole & $\ell_{\rm min}$ & 10\\
		Maximum multipole & $\ell_{\rm max}$ & \\
		-- Pessimistic & & $1500$\\
		-- Optimistic & & $5000$\\
        \hline
		\hline
        \multicolumn{3}{c}{$\GCph$}\\
        \hline
		Number of photo-$z$ bins & $N_z$ & 10 \\
		Galaxy number density & $\bar n_{\rm gal}$  & $30\,\mathrm{arcmin}^{-2}$ \\
		Minimum multipole & $\ell_{\rm min}$ & 10 \\
		Maximum multipole & $\ell_{\rm max}$ & \\
		-- Pessimistic & & $750$\\
		-- Optimistic & & $3000$\\
		\hline
		\hline
		\multicolumn{3}{c}{$\GCsp$}\\
		\hline
		Number of spectro-$z$ bins & $n_z$ & 4 \\
		Centres of the bins &$z_i$ & $\{1.0,\,1.2,\,1.4,\,1.65\}$\\
		Error on redshift & $\sigma_{0,z}$ & 0.001 \\
		Minimum scale & $k_{\rm min}$ & $0.001\,h\,{\rm Mpc}^{-1}$\\
		Maximum scale & $k_{\rm max}$ & \\
		-- Quasi-linear & & $0.15\,h\,{\rm Mpc}^{-1}$\\
		-- Pessimistic & & $0.25\,h\,{\rm Mpc}^{-1}$\\
		-- Optimistic & & $0.30\,h\,{\rm Mpc}^{-1}$\\
		\hline 
	\end{tabularx}
\end{table}

In this work, as in \citetalias{Blanchard:2019oqi}, we show the results for most of these single probes, but also for their combinations. It is important to mention that when we consider the combination of $\GCph$ with WL, we neglect any cross-correlation. However, when we add their cross-correlation \XCph, we include it both in the data vector and in the covariance, i.e.\ we perform a full analysis taking into account the cross-covariances between $\GCph$, WL, and their cross-correlation. This combination of three distinct two-point correlation functions is also known in the literature as 3x2pt and we will use this terminology interchangeably in this work. 
Moreover, again following \citetalias{Blanchard:2019oqi}, we do not present the values for $\GCph$ alone. The main reason for this choice is that we consider both $\sigma_8$ and the galaxy biases as parameters, which are degenerate in the linear regime. Even if this degeneracy might be partially broken when adding non-linear information, the Fisher formalism can still manifest numerical instabilities for this single probe alone. Therefore, we always show the constraints from \GCph\ in combination with other probes. In the optimistic scenario, we assume \GCsp\ to be uncorrelated to photometric probes. In the pessimistic setting, we neglect any correlation between \GCsp\ and WL, and also apply a redshift cut at $z<0.9$ for \GCph\ and \XCph, in order to minimise the overlap between the different galaxy clustering probes. Note however, that this redshift cut is only applied when combining spectroscopic and photometric data. Even in the pessimistic case, we do not apply any redshift cut when considering photometric data alone.

Finally, in
this work we have re-validated the codes used in \citetalias{Blanchard:2019oqi} 
to account for the modified recipe outlined in the previous sections. We  
have compared, for the constraint on each parameter of $\bm\Theta$ and the nuisance parameters of each probe, the performance of each independent
code to the median of the constraints obtained by the available codes. 
We have verified that the relative difference in percentage between a given code and such a median 
is always below 10\,\%, both for the marginalized and unmarginalized parameters, which was the threshold for code consistency adopted in \citetalias{Blanchard:2019oqi} and more recently in \cite{Euclid:2023pxu}.

\begin{table*}[htbp]
\caption{Forecast $1\sigma$ marginal relative errors on the cosmological parameters 
        for a flat $f(R)$ model with 
		$|f_{R0}|=5\times10^{-6}$ ($\logfr=-5.301$) 
		in the pessimistic and optimistic cases, using 
		Euclid observations of spectroscopic Galaxy Clustering 
		($\GCsp$), Weak Lensing (WL), photometric Galaxy Clustering 
		($\GCph$) and the cross-correlation among the photometric probes \XCph.}
		\begin{tabularx}{\textwidth}{Xcccccc}
			\hline
			\rowcolor{jonquil} \multicolumn{7}{c} {\boldsymbol{$|f_{R0}|=5\times10^{-6}$}} \\ 
			& \multicolumn{1}{c}{$\Omega_{\rm m,0}$} & \multicolumn{1}{c}{$\Omega_{\rm b,0}$} & \multicolumn{1}{c}{$\logfr$} & \multicolumn{1}{c}{$h$} & \multicolumn{1}{c}{$n_{\rm s}$} & \multicolumn{1}{c}{$\sigma_8$} \\
			\hline
			\rowcolor{cosmiclatte} Fiducial values & $0.32$ & $0.05$ & $-5.301$ & $0.67$ & $0.96$ & $0.853$ \\
			\hline
			\rowcolor{lavender(web)}\multicolumn{7}{l}{Pessimistic setting} \\ 
			\GCsp $(k_{\rm max} = 0.15\,h\,{\rm Mpc}^{-1})$   & $2.3\%$ & $4.6\%$ & $6.3\%$ & $0.6\%$ & $1.7\%$ & $1.9\%$  \\
			\GCsp $(k_{\rm max} = 0.25\,h\,{\rm Mpc}^{-1})$   & $1.4\%$ & $2.6\%$ & $3.6\%$ & $0.3\%$ & $1.1\%$ & $1.1\%$  \\
			WL                                                 & $2.3\%$ & $47\,\,\%$ & $8.3\%$ & $21\,\,\%$ & $4.8\%$ & $1.5\%$   \\
			\GCph                                    & $2.4\%$ & $5.8\%$ & $9.4\%$ & $4.1\%$ & $3.4\%$ & $1.9\%$ \\
			\GCsp+WL                                 & $0.7\%$ & $1.7\%$ & $2.3\%$ & $0.2\%$ & $0.8\%$ & $0.3\%$  \\
			\GCph+WL                                 & $1.0\%$ & $5.1\%$ & $3.8\%$ & $3.4\%$ & $1.9\%$ & $0.5\%$ \\
			\GCsp+WL+\GCph             & $0.6\%$ & $1.6\%$ & $2.2\%$ & $0.2\%$ & $0.7\%$ & $0.3\%$  \\
			WL+\GCph+\XCph             & $0.8\%$ & $5.0\%$ & $2.7\%$ & $3.3\%$ & $1.7\%$ & $0.4\%$ \\
			\GCsp+WL+\GCph+\XCph       & $0.6\%$ & $1.6\%$ & $1.7\%$ & $0.2\%$ & $0.7\%$ & $0.3\%$  \\
			\hline
			\rowcolor{lavender(web)}\multicolumn{7}{l}{Optimistic setting}  \\ 
			\GCsp $(k_{\rm max} = 0.30\,h\,{\rm Mpc}^{-1})$        & $1.3\%$ & $2.3\%$ & $3.0\%$ & $0.3\%$ & $1.0\%$ & $1.0\%$   \\
			WL                    									& $1.5\%$ & $45\,\,\%$ & $4.7\%$ & $19\,\,\%$ & $3.2\%$ & $0.9\%$           \\
			\GCph                    									& $1.5\%$ & $4.4\%$ & $1.7\%$ & $2.9\%$ & $0.7\%$ & $0.2\%$     \\
			\GCsp+WL            									& $0.5\%$ & $1.5\%$ & $2.0\%$ & $0.2\%$ & $0.6\%$ & $0.3\%$    \\
			\GCph+WL            									& $0.4\%$ & $4.3\%$ & $1.6\%$ & $2.1\%$ & $0.7\%$ & $0.2\%$     \\
			\GCsp+WL+\GCph    										& $0.3\%$ & $1.2\%$ & $1.2\%$ & $0.1\%$ & $0.3\%$ & $0.2\%$              \\
			WL+\GCph+\XCph         									& $0.3\%$ & $4.3\%$ & $1.4\%$ & $2.1\%$ & $0.7\%$ & $0.2\%$  \\
			\GCsp+WL+\GCph+\XCph 										& $0.3\%$ & $1.2\%$ & $1.0\%$ & $0.1\%$ & $0.3\%$ & $0.2\%$  \\
			\hline
		\end{tabularx}
	
		\label{tab:rel-errors-fR0-HS6}
	\end{table*}

\section{Results} \label{sec:results}

\begin{figure*}[htbp]
	\centering
	\includegraphics[width=0.75\linewidth]{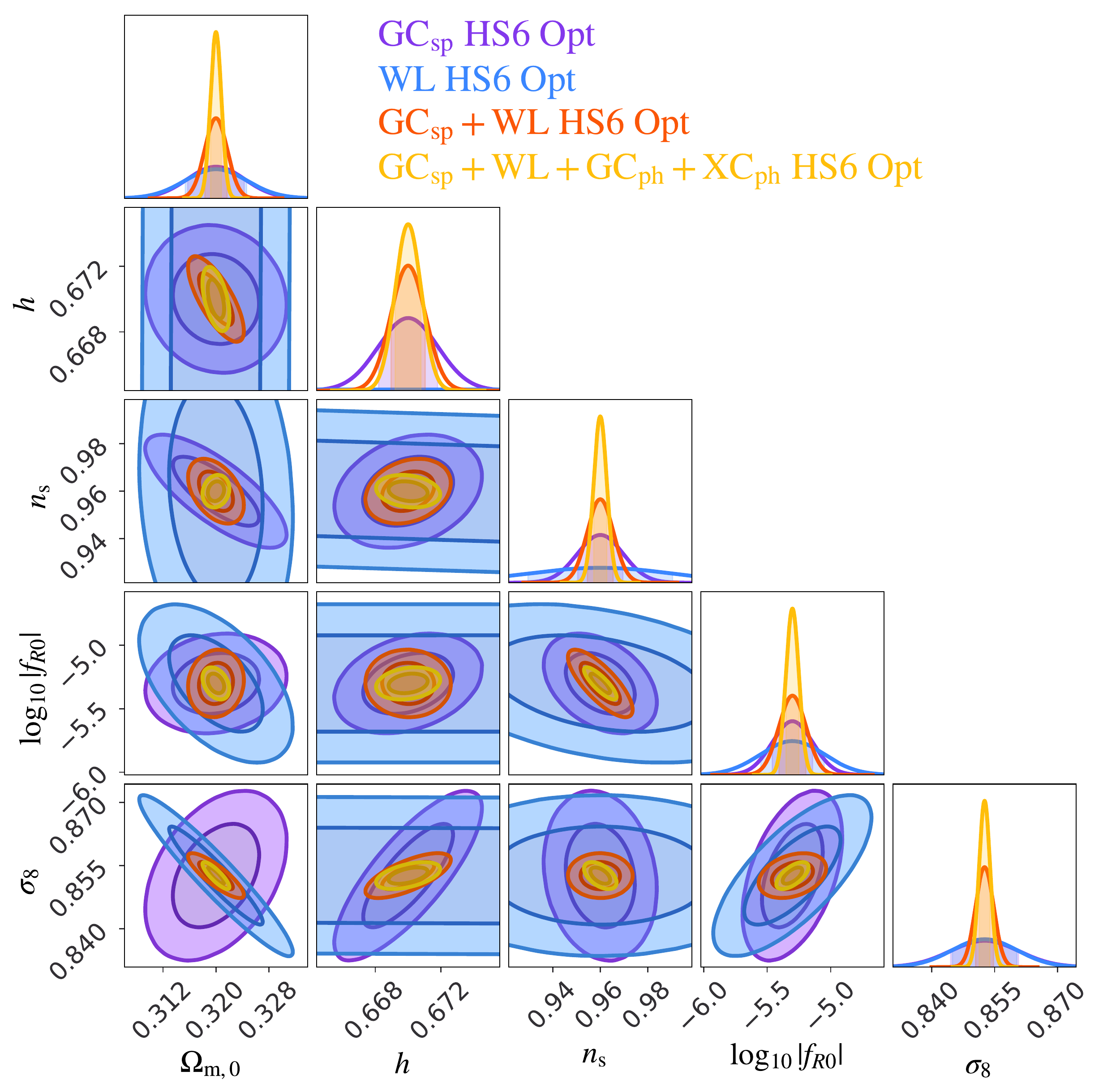}
	\caption{
	$1$ and $2\sigma$ joint marginal error contours on the cosmological parameters for a flat $f(R)$ model
	with $\fr=5\times10^{-6}$ in the optimistic scenario. Purple is for \GCsp, blue for WL, orange for the combination \GCsp+WL, and yellow 
	for all the photometric probes including their cross-correlation \XCph, combined with \GCsp, namely \GCsp+WL+\GCph+\XCph. 
	While the WL probe is unable to properly constrain the Hubble parameter $h$ and the primordial slope $n_{\rm s}$ 
	on its own, the orthogonality of the contours for \GCsp\ and WL in the subspaces involving 
	$h$ and $n_{\rm s}$, helps
	 to lift degeneracies and further improves the fully-marginalized constraints on $\logfr$,
	 when probe combinations are used.
	}
	\label{fig:ellipses-hs6-opt}
\end{figure*}

\subsection*{Optimistic scenario}
As shown in \cref{eq:fiducial-params},  we have chosen three different fiducial values of the Hu-Sawicki $f(R)$ model parameter, namely $|f_{R0}|=5\times10^{-7}$ (HS7), $|f_{R0}|=5\times10^{-6}$ (HS6) and $|f_{R0}|=5\times10^{-5}$ (HS5), based on our discussion of the current observational constraints in \cref{sec:intro}. We employ HS7, which contains a very small value of $\fr$ as our GR-limit test, since we cannot correctly perform forecasts at lower values of $\fr$ due to the limitations we have in the non-linear modelling with the Winther fitting formula mentioned in \cref{sec:non-linear}.
We remind the reader of the two considered scenarios, pessimistic and optimistic, as explained in \cref{sec:fisher} plus the `quasi-linear' scenario for \GCsp, defined by $k_{\rm max} = 0.15\,h\,{\rm Mpc}^{-1}$.

For our baseline fiducial (HS6) and in an optimistic setting, \Euclid alone will be able to constrain the additional parameter $\logfr$, which has a value of $\logfr=-5.301$ at the $1\sigma$ level with an absolute error of
\begin{itemize}
    \item $\sigma_{\logfr}=0.16$  with spectroscopic \GCsp\ alone \\ (corresponding to a relative $3.0\%$ error);\\
    \item $\sigma_{\logfr}=0.25$ with WL alone \\ (corresponding to a relative $4.7\%$ error);\\
    \item $\sigma_{\logfr}=0.07$ \\ combining WL, \GCph, and \XCph \\ (corresponding to a relative $1.4\%$ error);\\
    \item $\sigma_{\logfr}=0.05$  \\ using the full combination \GCsp+WL+\GCph+\XCph \\ (corresponding to a relative $1.0\%$ error).
\end{itemize}

\begin{figure}[htbp]
	\centering
	\includegraphics[width=0.9\linewidth]{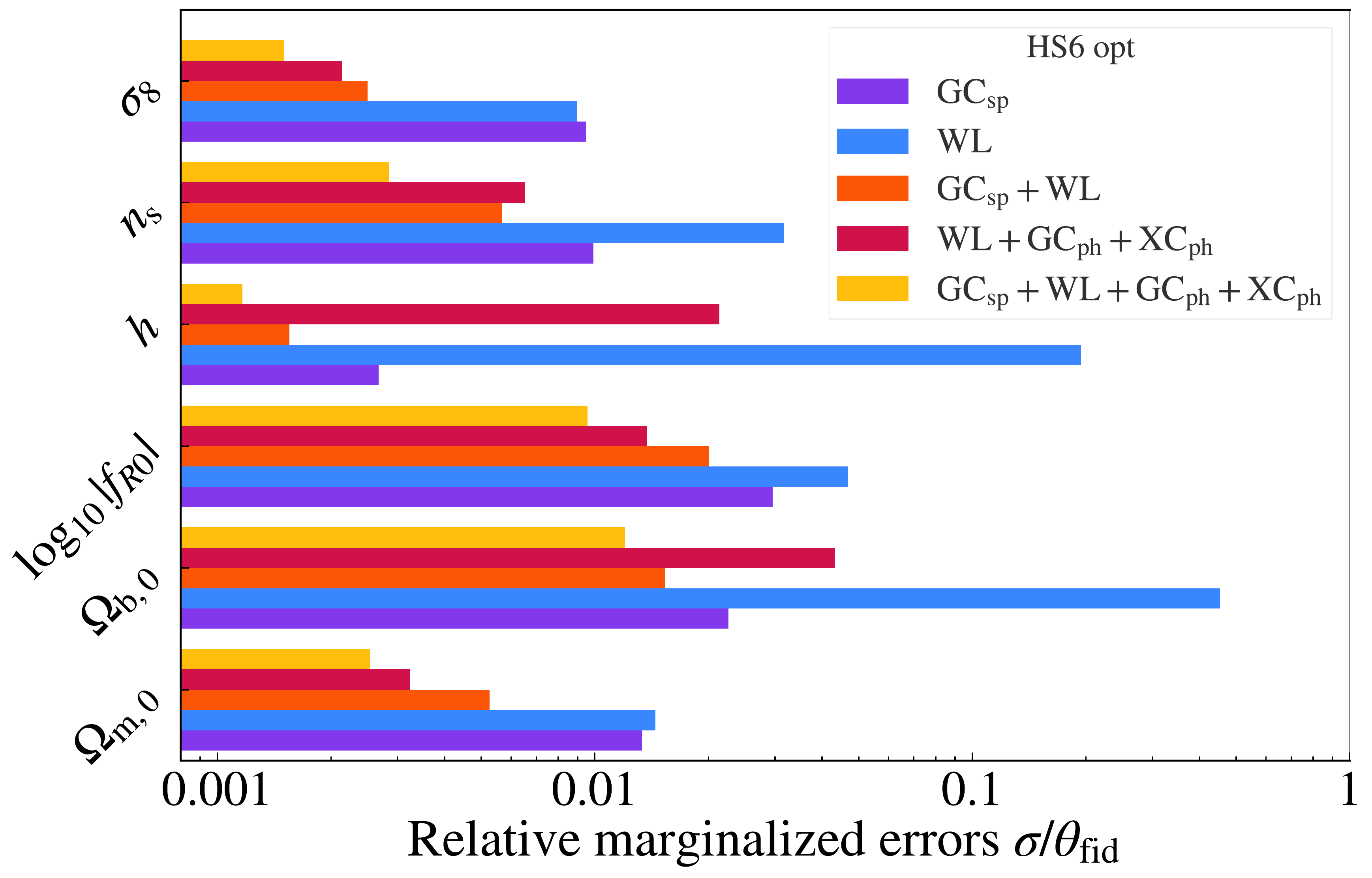}
	\caption{Marginalised $1\sigma$ errors on cosmological parameters, relative to their corresponding fiducial value for the optimistic case. Corresponding to the value of $\fr = 5\times 10^{-6}$. We show results for \GCsp\ (purple), WL (blue), \GCsp+WL (orange), the combination of all photometric probes, including cross correlations, WL+\GCph+\XCph\ (red) and the combination of all spectroscopic and photometric probes \GCsp+WL+\GCph+\XCph\ (yellow).
	We can see that the constraints on $\logfr$ and $\sigma_8$ are very similar among \GCsp\ and WL, but for other parameters like the Hubble parameter $h$ or the fraction of baryons $\Omega_{\rm b,0}$ the constraints coming from \GCsp\ alone are more stringent than the constraints of all photometric probes combined on their own.
	}
	\label{fig:barplot-hs6-opt}
\end{figure}

In \cref{tab:rel-errors-fR0-HS6} and in \cref{fig:barplot-hs6-opt}, 
we list the forecasted $1\sigma$ fully-marginalized errors (relative errors to its fiducial) on all the cosmological parameters considered for our model with $\fr = 5\times 10^{-6}$ (HS6), for the individual probes and their combinations in the optimistic scenario.
The probes shown are $\GCsp$ (in purple), WL (in blue), $\GCsp$+WL (orange), the 3x2pt combination of all photometric probes, including cross correlations, WL+\GCph+\XCph\ (red) and the combination of all spectroscopic and photometric probes \GCsp+WL+\GCph+\XCph\ (yellow). We keep the same color convention in all figures of the paper, when showing constraints from different probes.  
\Cref{tab:rel-errors-fR0-HS6} contains this information also for the pessimistic survey settings.

In \cref{fig:ellipses-hs6-opt}, we plot the elliptical $1\sigma$ and $2\sigma$ contours for the probes $\GCsp$, WL, the combination $\GCsp$+WL, and all the \Euclid probes combined \GCsp+WL+\GCph+\XCph, using the same color convention mentioned before. From the parameters used in the Fisher matrix, we leave out of this plot $\Omegab$ since it does not provide any extra insight.
As it can be seen from this figure, \GCsp\ is always better at constraining $h$ and $n_{\rm s}$, compared to the cosmic shear probe (WL) on its own. 
However, due to the orthogonality of the contours, especially in the subspaces combining $\sigma_8$ with $\Omegab$, $n_{\rm s}$ and $h$, there is an important lifting of degeneracies, which makes the combination of \GCsp\ and WL (shown in orange) much more constraining. The relative constraint on $\logfr$ coming from WL alone (blue) is of the order of $4.7\%$, this reduces by more than a factor of two when combining with the spectroscopic probe (orange) and another factor two when adding all \Euclid probes together (yellow), yielding in total a relative constraint on $\logfr$ of the order of $1.0\%$ in this optimistic case.
Throughout this work, we indicate the $1\sigma$ constraints rounded to the nearest significant digit, since our Fisher matrix method has been validated at the 10\% level on the discrepancy between $1\sigma$ marginalized and unmarginalized errors on the cosmological and nuisance parameters, as mentioned in \cref{sec:fisher}.

Contour ellipses comparing the constraints from the combination of spectroscopic GC and WL to the 3x2pt photometric combination (WL+$\GCph$+$\XCph$) can be found in \cref{fig:ellipses-A3} in the Appendix (using the same global color convention), where it can be seen that it is the spectroscopic probe that helps to break degeneracies in the $h$ and $\Omegab$ planes, mainly due to the sensitivity of the BAO wiggles on these two parameters. 
The cosmic shear probe (WL) is relatively insensitive to $\Omegab$ and $h$ and the full combination of photometric probes is also not good at constraining $h$. It is the breaking of degeneracies when combining \GCsp\ and WL probes, that improves considerably the constraints on all parameters, showcasing the power of \Euclid to measure parameters
in and beyond the standard model of cosmology.

\subsection*{Constraints on the fundamental model parameter $\fr$}
Note that we perform the Fisher matrix analysis on the parameter $\logfr$, instead of directly on $\fr$, since for very small numbers and for large order of magnitude differences, the Fisher matrix derivatives might become unstable \citep[see, e.g., ][Appendix~A1]{2018MNRAS.481.1251C}. Therefore, it is recommended to have all the involved parameters in the Fisher matrix being of the same order of magnitude. 
Since the transformation between $\logfr$ and $\fr$ is non-linear and the parameter constraints are not small in some cases,
we cannot simply use a Jacobian transformation to convert between the Fisher matrices in this case. 
Our assumption of Gaussianity is only true for the logarithmic parametrization $\logfr$, therefore the posterior contours for 
$\fr$ will be non-Gaussian.

We can, nevertheless, obtain the fully marginalized constraints on $\fr$ by transforming the $\logfr$ symmetric bounds 
\begin{equation}\label{eq:error-prop}
    \logfr^{(\pm)} = \logfr_{\rm fid} \pm \sigma_{\logfr},
\end{equation}
into the upper and lower bounds for the linearly parameterized parameter $\fr$
\begin{equation}\label{eq:error-fr0}
    {\fr}^{(\pm)} = 10^{\left(\logfr_{\rm fid} \pm \sigma_{\logfr}\right)} = \fr_{\rm fid}\times 10^{\left(\pm\sigma_{\logfr}\right)}\,.
\end{equation}
This will yield asymmetrical errors in $\fr$ since the upper and lower bounds will be given by the exponentiation of the symmetrical $1\sigma$ bounds.

Using these formulas we can obtain the upper and lower $1\sigma$-bounds for our fiducial parameter $\fr$ for the different cases

\begin{itemize}
    \item $\fr=(5.0^{+ 2.2}_{-1.5} \times 10^{-6})$  with spectroscopic \GCsp\ alone;\\
    \item $\fr=(5.0^{+ 3.9}_{-2.2} \times 10^{-6})$ with WL alone;\\
    \item $\fr=(5.0^{+ 0.91}_{-0.77} \times 10^{-6})$ combining WL, \GCph, and \XCph \\
    \item $\fr=(5.0^{+ 0.62}_{-0.55} \times 10^{-6})$  \\with the combination \GCsp+WL+\GCph+\XCph.
\end{itemize}

As one can clearly see from these numbers, the stronger the constraint on the $\logfr$ parameter, the more symmetric the upper and lower bounds on $\fr$ become, simply due to the central limit theorem and the fact that for a very peaked likelihood, a Gaussian approximation is always possible around the maximum of the posterior distribution.

In order to visualize this, we take our Fisher matrices computed for each of the cases and probes and assign them to a multivariate
Gaussian distribution, by using the Fisher matrix as the inverse covariance.
We then sample from this distribution and transform the samples of $\logfr$ into samples on $\fr$ using the inverse of the logarithmic transformation, see \cref{eq:error-fr0}.
Using this technique, we plot on the left side of \cref{fig:1d-contours-fr0} the marginalized posterior 1-d and 2-d distributions for the original model parameter $\fr$ in the HS6 case, for the optimistic scenario, using the same color convention as before.
Compared to \cref{fig:ellipses-hs6-opt} we show now, for illustration purposes, the 3x2pt combination of photometric probes
(WL+\GCph+\XCph), instead of the $\GCsp$+WL combination, and we focus only on the parameters $\fr$, $\Omegam$ and $\sigma_8$.
For the subspaces involving the $\fr$ parameter, we see now non-symmetrical posterior distributions, which are clearly non-Gaussian, and noticing the marginalized 1-d posterior distributions in the top of each sub-panel, we see that they have a longer tail on the upper side of the constraints. 
This asymmetry is also due to the fact that using the logarithmic parameterisation automatically prevents us from going into negative regions of the parameter space for $\fr$, therefore the probability distributions are skewed towards the positive side.
However, the stronger the constraints on this parameter (for example in the case of the full combination of the \Euclid probes, in yellow), the better the Gaussian approximation.

\begin{figure}[htbp]
	\centering
	\includegraphics[width=0.95\linewidth]{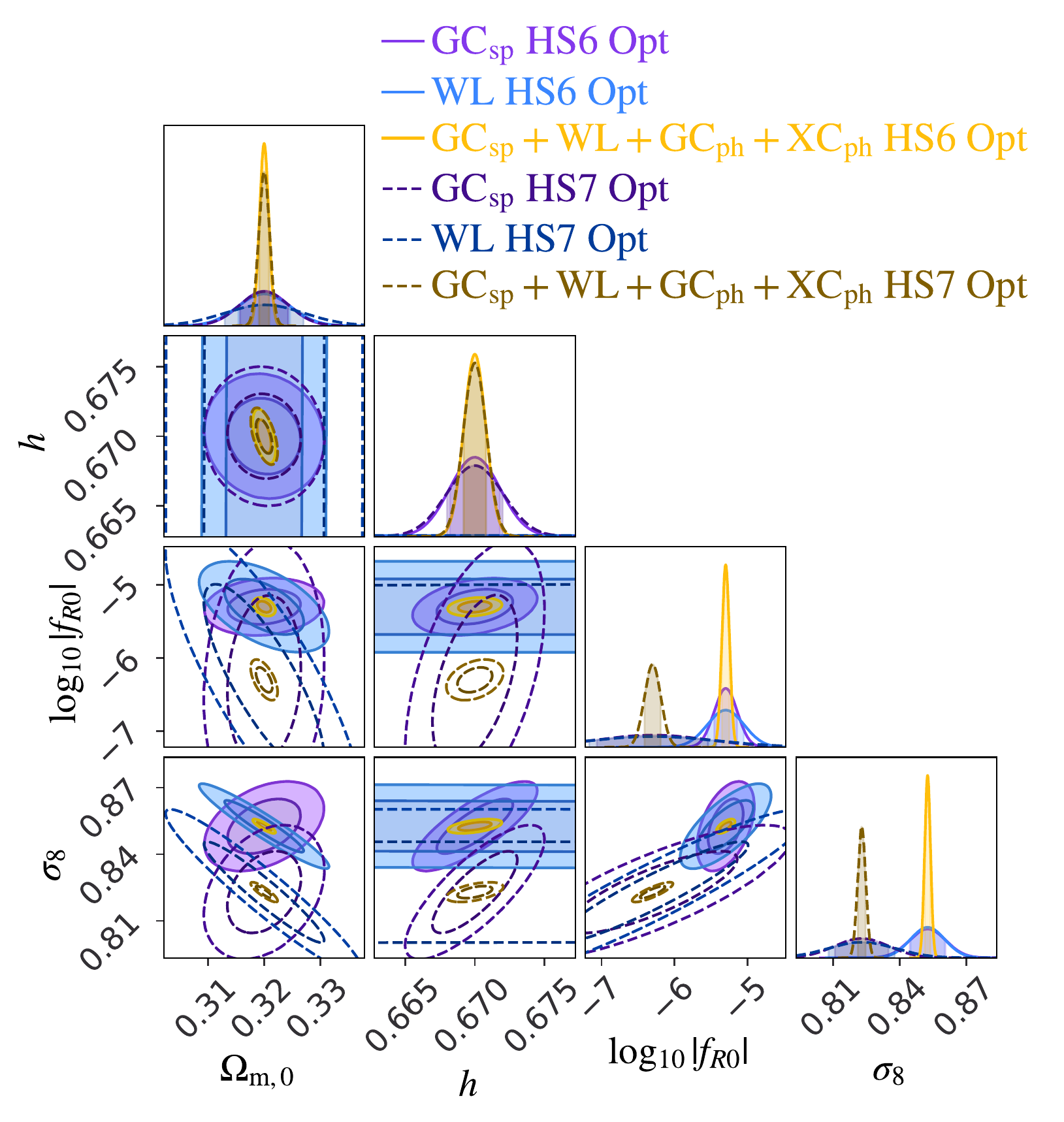}
    \caption{
	1 and 2$\sigma$ joint marginal error contours on the cosmological parameters for a flat $f(R)$ model with $\fr=5\times10^{-6}$ (HS6) vs. $\fr=5\times10^{-7}$ (HS7) in the optimistic scenario. 
	Lighter colors and solid contours correspond to HS6, while darker colors and dashed empty contours correspond to HS7.
	In purple the spectroscopic \GCsp, in blue the WL probe, in yellow the photometric and spectroscopic probes 
	combined together (\GCsp+WL+\GCph+\XCph).
	One can see that for certain parameter combinations like $\Omegam$ vs $h$ the contours are very similar, in size and orientation, since the fiducial values for the standard cosmological parameters in these models only differ by their value of $\sigma_8$. 
	However, for $\logfr$ in HS7 the ellipses become much more elongated, due to this model being very close to $\lcdm$ and therefore, being less  
	constrained in its extra model parameter compared to HS6.
	}
	\label{fig:ellipses-comparison}
\end{figure}

\subsection*{Comparison of the HS5, HS6 and HS7 Hu-Sawicki models}
\begin{figure*}[htbp]
	\centering
	\includegraphics[width=0.49\linewidth]{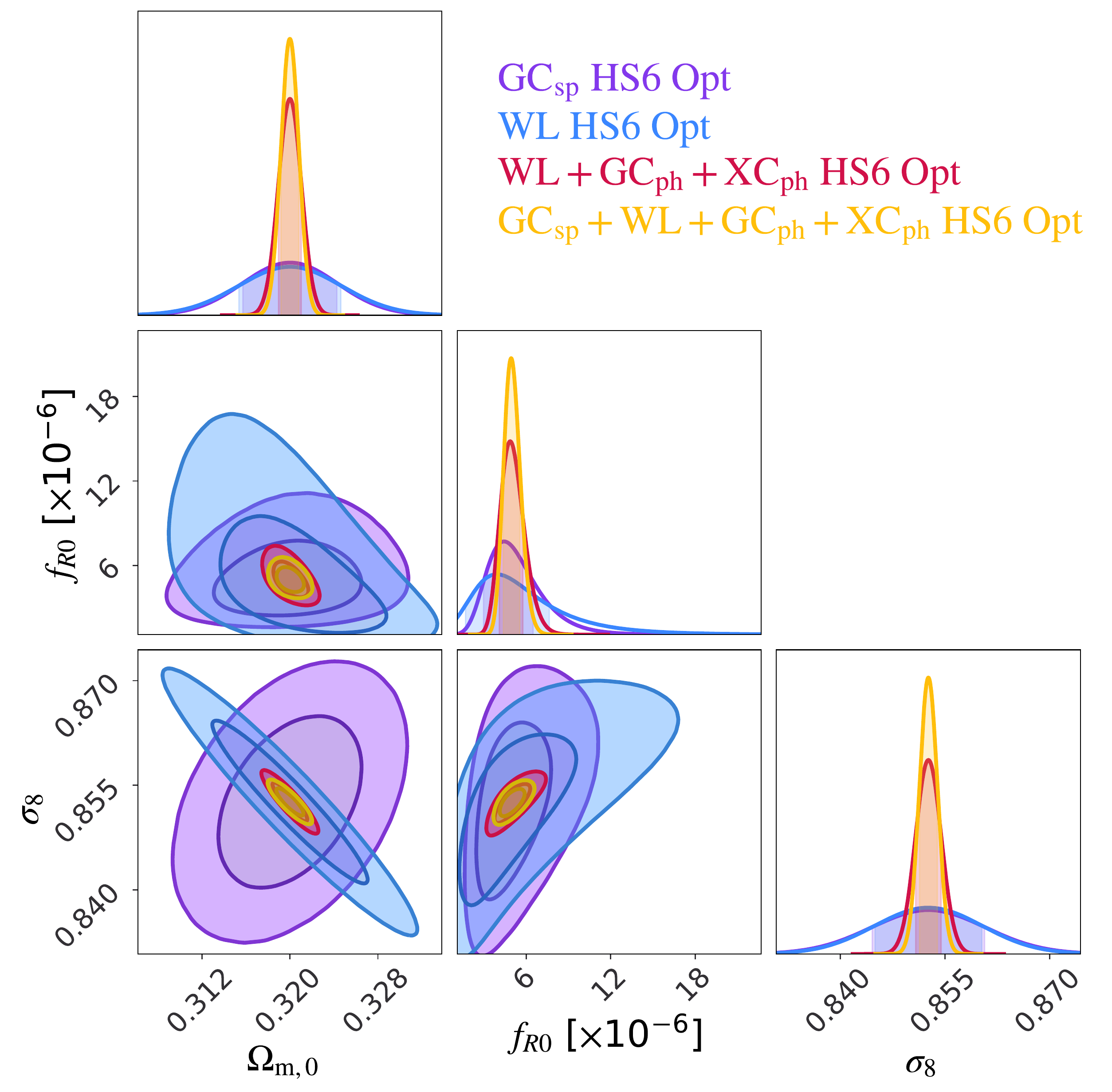}
	\includegraphics[width=0.49\linewidth]{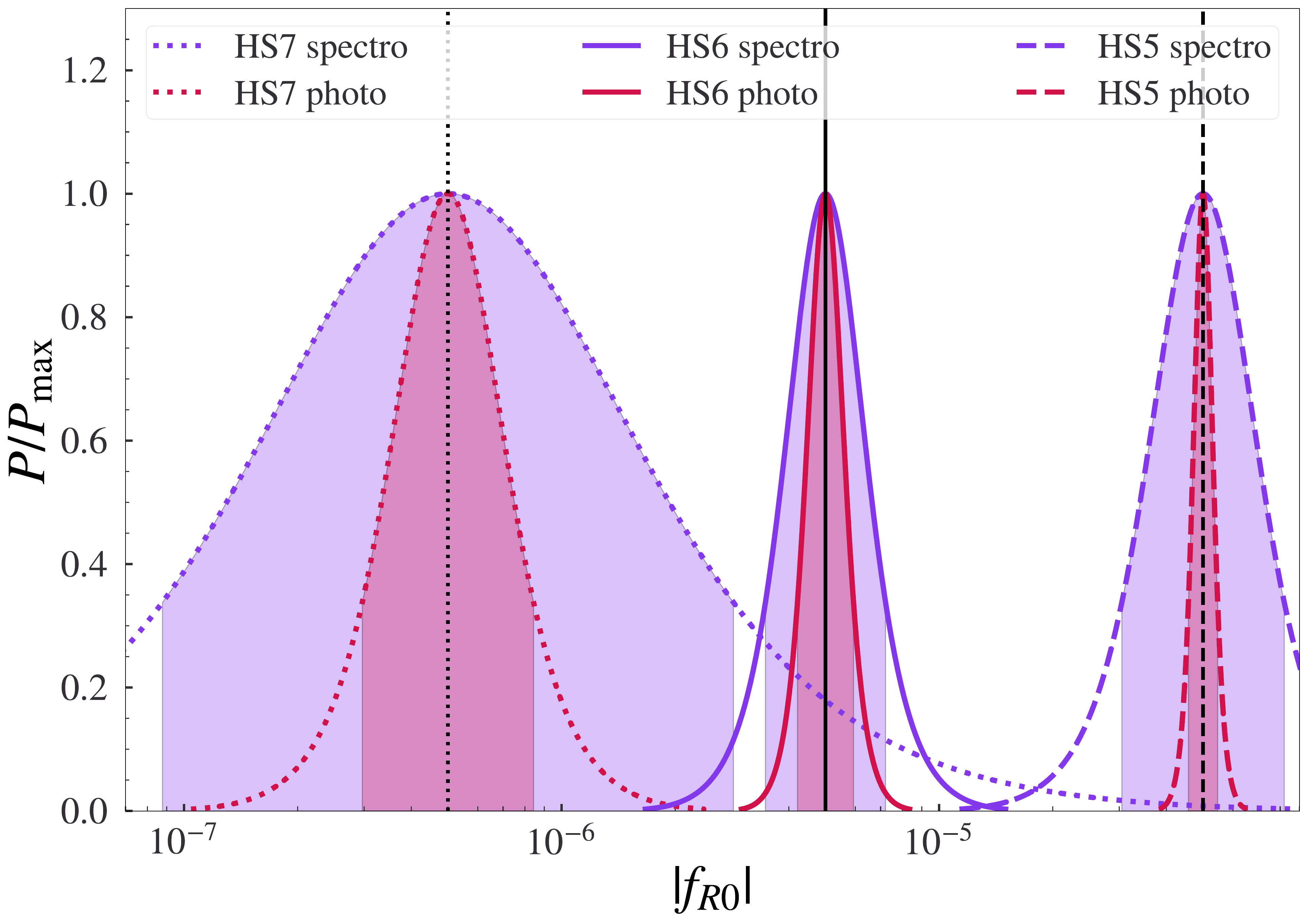}
	\caption{
    \textbf{Left:} Same as \cref{fig:ellipses-hs6-opt} (HS6) for a subset of the cosmological
	parameters when transforming the contours back into the fundamental parameter $\fr$. 
	While in the logarithmic case the marginalized Fisher contours 
	are simply Gaussians by definition, 
	this is not the case for the contours on the linearly parameterized case $\fr$.
	For the probes $\GCsp$ (purple) and WL (blue) on their own which are less constraining, 
	the errors on $\fr$ are very unsymmetric with heavy tails on the right. 
	However, the smaller the  $1\sigma$ marginalized error on $\logfr$ is, the
	better can the marginalized posterior on $\fr$ be approximated by a Gaussian.
	This can be well seen for the full combination of \Euclid probes (yellow).
	\textbf{Right: }
	1-dimensional fully-marginalized posteriors on the $\fr$ parameter for 
	$\fr = 5\times 10^{-7}$ (HS7, dotted lines), 
	$\fr = 5\times 10^{-6}$ (HS6, solid lines) and $\fr = 5\times 10^{-5}$ (HS5, dashed lines), 
	after applying the transformation
	in \cref{eq:error-fr0}. Here we show the spectroscopic probe $\GCsp$ on its own in purple,
	and the 3x2pt combination of photometric probes (WL+$\GCph$+$\XCph$) in red, using the same color
	convention throughout the whole paper.
	The shaded areas correspond to the $1\sigma$ confidence intervals, while the lines extend up to
	$3\sigma$. We can see that all three models can be differentiated at $1\sigma$ from each other,
	even using the spectroscopic probe on its own. 
	However, the larger tail of HS7 on the right side, makes it impossible to discriminate this model 
	from HS6 at $2\sigma$ and from HS5 at $3\sigma$.
	Using the full combination of photometric probes, all models can be distinguished at more than 
	$3\sigma$ from each other.
	}
	\label{fig:1d-contours-fr0}
\end{figure*}

Besides our baseline fiducial for $f(R)$ gravity, $\fr=5\times 10^{-6}$, we also perform forecasts on a second and third fiducial, namely $\fr=5\times 10^{-5}$ (HS5) and $\fr=5\times 10^{-7}$ (HS7), the latter  is closer to \lcdm\ in terms of perturbations and background evolution. 
A closer limit to \lcdm\ is currently not possible using non-linear predictions, since our non-linear fitting formula discussed in \cref{sec:non-linear} cannot be extrapolated for values of the $f(R)$ model parameter smaller than $\fr=5\times 10^{-7}$.
We also perform forecasts on a flat \lcdm\ model in order to have a comparison scenario within the assumptions of our recipe for the large scale structure probes. 

In \cref{tab:rel-errors-fR0-HS5-HS7} we report the $1\sigma$ fully marginalized forecasted errors for these models
in the optimistic and pessimistic scenarios for the most important \Euclid probe-combinations discussed above.
For the HS5 model we find in general very similar constraints to HS6, but it is important to remark that due to the increased signal at non-linear scales and also the larger $k$-dependence of the growth, due to a higher value of $\fr$, we obtain better constraints on $\logfr$ for the full combination of probes, both in the pessimistic and optimistic scenarios, with relative errors of $1.4\%$ and $0.6\%$, respectively. 
In the case of the HS7 model, the fiducial value of $\fr$ is yet another order of magnitude closer to \lcdm\ as compared to HS6 and therefore, we expect the extra signal on the power spectrum coming from modified gravity to be much smaller. 
We indeed find that this is the case, and compared to HS6, the relative percentage constraints on $\logfr$ increase by a factor 4.8 in \GCsp, a factor 3.5 in WL and a factor 2.2 when considering the full \Euclid spectroscopic and photometric combination. The final relative constraint on $\logfr$ for this latter case is $1.8\%$, which corresponds to a determination of $\fr=(5.0^{+ 1.5}_{-1.1} \times 10^{-7})$. 
In \cref{fig:ellipses-comparison} we plot the confidence contours for this model HS7 in the optimistic scenario compared to the HS6 baseline model. In dark blue/blue we plot the constraints for \GCsp\ for HS7/HS6, in dark purple/violet the constraints on WL for HS7/HS6 and in yellow/orange the constraints on the full spectroscopic+photometric \Euclid combination. 
For the standard cosmological parameters, such as in the subspace of 
$\Omegam$ vs. $h$ or $\sigma_8$, we can see that we recover very similar contours in size and in orientation.
On the other hand, one can see in the panel of $\logfr$ vs. $\sigma_8$ that the \GCsp\ and WL probes alone are not sufficient to distinguish these two models from each other, but the full spectroscopic+photometric combination is able to discriminate between them at several $\sigma$.
As discussed previously, this is under the assumption that both models have the same primordial amplitude and, therefore, due to a larger growth in the case of HS6, they end up with different values of $\sigma_8$ today.
To be certain that the choice of $\sigma_8$ values does not affect our constraints, we performed a Fisher matrix forecast of the HS6 model, changing its primordial amplitude $A_{\rm s}$, so that it has the same amplitude of fluctuations $\sigma_8=0.816$ today as the baseline $\lcdm$ model. 
We found that none of the parameter constraints changed considerably with respect to the baseline HS6 case.

We have also tested that when fixing the extra model parameter $\logfr$ in the HS7 model, which is the closest we have to $\lcdm$ and just deviates in the non-linear power spectrum and the growth rate function by a few percent (see \cref{fig:fr-scale-dep,fig:fr-fitting-formula}), we then recover the same constraints on the five remaining $\lcdm$ parameters. This comparison is shown in the right panel of \cref{fig:ellipses-A3}, where in dark yellow
we show the full combination of \Euclid probes for the pessimistic HS7 case and in cyan the corresponding 1-d and 2-d contours for the $\lcdm$ scenario as described in \citetalias{Blanchard:2019oqi}.

On the right panel of \cref{fig:1d-contours-fr0}, we compare now the constraints across the three models studied in this work,
HS5 (in dashed lines), HS6 (in solid lines) and HS7 (in dotted lines) for the fundamental parameter $\fr$. 
We plot the 1-d fully-marginalized posterior on $\fr$ for the spectroscopic probe $\GCsp$ on its own and the 3x2pt photometric combination of probes, using again the same color codes as before. The shaded areas under the curve represent the $1\sigma$ confindence intervals and the extent of the lines corresponds to a $3\sigma$ deviation from the mean.
Due to the logarithmic scale of the $x$-axis, the distributions look symmetric again, nevertheless we can see that in absolute $\fr$ numbers, the constraints are much larger on the right side of the constraints.
In the HS7 model for the optimistic scenario, the forecasted upper bound on $\fr$ would be $2.9 \times 10^{-6}$ at $1\sigma$, $1.7 \times 10^{-5}$ at $2\sigma$ and $9.6 \times 10^{-5}$ at $3\sigma$.
For instance, assuming data would follow the HS7 model, using only the spectroscopic probe, it would not be possible to distinguish it from HS6 at the $2\sigma$ confidence level. Discriminating HS7 from HS5 would be impossible at a $3\sigma$ confidence level. Logically, this situation degrades even more in the pessimistic scenario.
However, when using the full combination of photometric probes (red lines and shaded regions), all three models would be distinguishable from each other at more than $3\sigma$.
The 1-, 2- and $3\sigma$ level upper bounds for HS7 are in this combination of probes of the order of $8.4 \times 10^{-7}$, $1.4 \times 10^{-6}$
and $2.4 \times 10^{-6}$, respectively.
Using \cref{eq:error-fr0}, the rest of the non-symmetrical bounds on $\fr$ can be recovered for all the numbers listed in \cref{tab:rel-errors-fR0-HS6,tab:rel-errors-fR0-HS5-HS7}. Since we performed the Fisher matrix analysis on the logarithm of the original model parameter, we opt to report here only the forecasted Fisher matrix constraints which are Gaussian and therefore, symmetric.

Under our given assumptions (and the fact that we have fixed for both models the sum of neutrino masses and the same initial amplitude of perturbations $A_{\rm s}$), we can say that the full combination of \Euclid probes would be capable of distinguishing between these three $f(R)$ models with a high degree of certainty.
Since most of the constraints in the photometric probe come from small scales, this highlights again the importance of using the deeply non-linear regime of structure formation to test this $f(R)$ theory accurately and in an ideal scenario, discern it from the $\lcdm$ model, provided the data indicates a strong preference for the standard cosmological model with a large statistical significance.

\subsection*{Pessimistic survey scenario}
For the baseline HS6 model and in the pessimistic scenario, the errors on $\logfr$, using \GCsp, do not degrade that much, increasing to $3.6\%$ when scales of $k_{\rm max} = 0.25\,h\,{\rm Mpc}^{-1}$ are considered, while increasing to $6.3\%$ when the cut is performed at a smaller wavenumber $k_{\rm max} = 0.15\,h\,{\rm Mpc}^{-1}$, which we label as the quasi-linear scenario in this work.
On the other hand, for photometric probes the pessimistic scenario has a much larger impact on the errors, because scales are decreased from $\ell_{\rm max}=5000$ to $1500$ in the case of WL and from $\ell_{\rm max}=3000$ to $750$ in the case of \GCph\ (see \cref{tab:specifications-ec-survey}). For WL alone, the errors degrade by a factor 1.7, for the combination of all photometric probes, the errors become 2 times larger, and since these probes dominate the total constraining power on this parameter, for the combination of spectroscopic and photometric probes, the errors on $\logfr$ are a factor of 1.7 larger than compared to the optimistic scenario.
For HS5 the degradation between the optimistic and pessimistic scenarios are of the same order, of about a factor 2, while for HS7 the degradation is stronger and it is of a factor 3, when photometric probes are involved.

Our tests demonstrate that the non-linear scales have a much larger impact on the photometric probes, since in order to probe multipoles of about $\ell=5000$, one needs to have an accurate determination of the power spectrum in the deeply non-linear regime, at least to scales of about $k=5\,h\,{\rm Mpc}^{-1}$ at $z=0$, before noise and other systematic effects, like baryonic contributions start dominating \citep[for an analysis of the effect of $k$-cuts on the information content in cosmic shear, see][]{Taylor:PhysRevD.98.083514}. 
In the right panel of \cref{fig:ellipses-A2}, we compare the Fisher matrix contours for the optimistic scenario (yellow) versus the pessimistic scenario (blue), within our baseline model and using all \Euclid primary probes combined.

\begin{table*}[htbp]
	\caption{Forecast $1\sigma$ marginal relative errors on the cosmological parameters for a flat $f(R)$ model with $|f_{R0}|=5\times10^{-5}$ , $|f_{R0}|=5\times10^{-7}$ and the flat $\lcdm$ model in the pessimistic and optimistic scenarios.}
	\begin{tabularx}{\textwidth}{Xcccccc}
		\hline
		\rowcolor{jonquil} \multicolumn{7} {c} {{{\boldsymbol{$|f_{R0}|=5\times10^{-5}$}}}} \\ 
		& \multicolumn{1}{c}{$\Omega_{\rm m,0}$} & \multicolumn{1}{c}{$\Omega_{\rm b,0}$} & \multicolumn{1}{c}{$\logfr$} & \multicolumn{1}{c}{$h$} & \multicolumn{1}{c}{$n_{\rm s}$} & \multicolumn{1}{c}{$\sigma_{8}$} \\
		\hline
		\rowcolor{cosmiclatte} Fiducial values & \multicolumn{1}{c}{$0.32$} & \multicolumn{1}{c}{$0.05$} & \multicolumn{1}{c}{$-4.301$} & \multicolumn{1}{c}{$0.67$} & \multicolumn{1}{c}{$0.96$} & \multicolumn{1}{c}{$0.911$} \\
		\hline
		\rowcolor{lavender(web)}\multicolumn{7} {l}{Pessimistic setting}  \\ 
		\GCsp   			&1.5\%  & 2.9\% & 5.2\% & 0.4\% & 1.3\% & 1.0\%   \\
		WL+\GCph+\XCph 		&0.8\%  & 4.9\% & 1.9\% & 2.8\% & 0.9\% & 0.5\%  \\
		\GCsp+WL+\GCph+\XCph 	&0.6\%  & 1.6\% & 1.4\% & 0.2\% & 0.5\% & 0.3\%  \\
		\hline
		\rowcolor{lavender(web)}\multicolumn{7} {l}{Optimistic setting}  \\ 
		\GCsp   			            &1.5\%  & 2.7\% & 5.0\% & 0.3\% & 1.2\% & 0.9\%   \\
		WL+\GCph+\XCph 		    &0.3\%  & 4.3\% & 0.9\% & 2.1\% & 0.5\% & 0.1\%  \\
		\GCsp+WL+\GCph+\XCph 	&0.2\%  & 1.2\% & 0.6\% & 0.1\% & 0.2\% & 0.1\%  \\
		\hline
		\hline
		\rowcolor{jonquil} \multicolumn{7} {c} {{{\boldsymbol{$|f_{R0}|=5\times10^{-7}$}}}} \\ 
		& \multicolumn{1}{c}{$\Omega_{\rm m,0}$} & \multicolumn{1}{c}{$\Omega_{\rm b,0}$} & \multicolumn{1}{c}{$\logfr$} & \multicolumn{1}{c}{$h$} & \multicolumn{1}{c}{$n_{\rm s}$} & \multicolumn{1}{c}{$\sigma_{8}$} \\
		\hline
		\rowcolor{cosmiclatte} Fiducial values & \multicolumn{1}{c}{$0.32$} & \multicolumn{1}{c}{$0.05$} & \multicolumn{1}{c}{$-6.301$} & \multicolumn{1}{c}{$0.67$} & \multicolumn{1}{c}{$0.96$} & \multicolumn{1}{c}{$0.823$} \\
		\hline
		\rowcolor{lavender(web)}\multicolumn{7} {l}{Pessimistic setting}  \\ 
		\GCsp   			            &1.4\%  & 2.8\% & 14\% & 0.4\% & 1.0\% & 1.8\%   \\
		WL+\GCph+\XCph 		    &0.8\%  & 5.4\% & 8.9\% & 4.0\% & 2.0\% & 0.7\%  \\
		\GCsp+WL+\GCph+\XCph 	&0.6\%  & 1.6\% & 5.4\% & 0.2\% & 0.6\% & 0.4\%  \\
		\hline
		\rowcolor{lavender(web)}\multicolumn{7} {l}{Optimistic setting}  \\ 
		\GCsp   			            &1.3\%  & 2.4\% & 12\% & 0.3\% & 0.9\% & 1.5\%   \\
		WL+\GCph+\XCph 		    &0.6\%  & 4.3\% & 3.6\% & 2.3\% & 0.7\% & 0.5\%  \\
		\GCsp+WL+\GCph+\XCph 	&0.3\%  & 1.3\% & 1.8\% & 0.1\% & 0.2\% & 0.2\%  \\
		\hline
		\hline
		\rowcolor{jonquil} \multicolumn{7} {c} {{{flat \boldsymbol{$\lcdm$}}}} \\ 
		& \multicolumn{1}{c}{$\Omega_{\rm m,0}$} & \multicolumn{1}{c}{$\Omega_{\rm b,0}$} & \multicolumn{1}{c}{-} & \multicolumn{1}{c}{$h$} & \multicolumn{1}{c}{$n_{\rm s}$} & \multicolumn{1}{c}{$\sigma_{8}$} \\
		\hline
		\rowcolor{cosmiclatte} Fiducial values & \multicolumn{1}{c}{$0.32$} & \multicolumn{1}{c}{$0.05$} & \multicolumn{1}{c}{-} & \multicolumn{1}{c}{$0.67$} & \multicolumn{1}{c}{$0.96$} & \multicolumn{1}{c}{$0.816$} \\
		\hline
		\rowcolor{lavender(web)}\multicolumn{7} {l}{Pessimistic setting}  \\ 
		\GCsp & 1.3\% & 2.1\% & - & 0.3\% & 0.9\% & 0.9\% \\
        WL+\GCph+\XCph & 0.8\% & 5.2\% & - & 2.7\% & 0.9\% & 0.4\% \\
        \GCsp+WL+\GCph+\XCph & 0.5\% & 1.3\% & - & 0.2\% & 0.5\% & 0.2\% \\
		\hline
		\rowcolor{lavender(web)}\multicolumn{7} {l}{Optimistic setting}  \\ 
		\GCsp & 1.2\% & 1.9\% & - & 0.2\% & 0.8\% & 0.8\% \\
        WL+\GCph+\XCph & 0.3\% & 4.6\% & - & 2.0\% & 0.4\% & 0.1\% \\
        \GCsp+WL+\GCph+\XCph & 0.2\% & 0.9\% & - & 0.1\% & 0.1\% & 0.1\% \\
		\hline
	\end{tabularx}
	\label{tab:rel-errors-fR0-HS5-HS7}
\end{table*}

\section{Conclusions} \label{sec:conclusions}

In this work we have focused on the ability of the future \Euclid mission to constrain extensions of the concordance cosmological model. The large literature on $f(R)$ cosmologies, as well as their properties (screening mechanism, scale-dependent growth of structures, possibility to rewrite it as a scalar-tensor theory embedded in the Horndeski action), make $f(R)$ an ideal test case of modified gravity models. In this work we have considered \citet{Hu:2007nk} $f(R)$ cosmologies and predicted forecasts on their peculiar parameter, $f_{R0}$, and other cosmological parameters.

As done in \citetalias{Blanchard:2019oqi}, forecasts are validated and computed for spectroscopic galaxy clustering (\GCsp), photometric galaxy clustering (\GCph), weak lensing cosmic shear (WL), and the cross-correlation between the last two probes (\XCph). The extension presented in this paper over the forecasting pipeline developed in \citetalias{Blanchard:2019oqi} is threefold: first, for the class of cosmologies investigated here, which goes beyond \lcdm; secondly, in the encoded equations for all probes, as they now include a growth factor which is scale dependent; and finally, in the software development of Fisher matrix codes, which are now validated also for this extended recipe. Validation was pursued as follows:
\begin{enumerate}
    \item We have first compared the input quantities obtained from different Boltzmann solvers and compared the quasi-static limit with the full evolution, finding equivalent results for the redshifts and scales of interest---in particular, predictions on the matter power spectrum up to $k = 10\,h\,\mathrm{Mpc}^{-1}$ lead to sub-percent agreement between quasi-static and full evolution.
    \item We have then used the forecasting pipeline presented in \citetalias{Blanchard:2019oqi} and modified it to account for the scale-dependent quantities and the other MG modifications arising in $f(R)$.
    \item As done in \citetalias{Blanchard:2019oqi}, we have implemented independent forecast codes and considered their output validated when the discrepancy on each parameter uncertainty, compared to the median, was less than $10\%$, both before and after marginalisation.
\end{enumerate}

Hence, we have computed forecasts in an optimistic and a pessimistic scenario, depending on the range in scales considered for each probe and on the level of systematics included. This choice follows the one in \citetalias{Blanchard:2019oqi}, for easier comparison, and it is summarised in \cref{tab:specifications-ec-survey}. In the specific case of \GCsp, we have also shown the impact of a quasi-linear, more conservative choice, due to the additional uncertainty in the non-linear predictions for a theoretical model beyond \lcdm.

In an optimistic scenario, combining all \Euclid primary probes, and considering our baseline fiducial $\fr=5\times10^{-6}$, we have obtained that \Euclid alone will be able to constrain $\logfr$ at the $1.0$\% level using the full combination \GCsp+WL+\GCph+\XCph. This uncertainty increases to $1.4$\%, when considering only the photometric probes, which highlights the importance of combining spectroscopic and photometric observables. The constraint considering the spectroscopic probe alone increases to $3\%$.
We have discussed in the text how these constraints on the logarithmic parameter transform into asymmetrical constraints of the original model parameter $\fr$.
The $\logfr$ parametrization has been used several times in the literature, not only for theoretical considerations but also in Markov-Chain-Monte-Carlo explorations of cosmological likelihoods, such as in \citet{Dossett:2014oia, Schneider:2019xpf} and \citet{KiDS:2020ghu}. 
However, for future analysis with real data, one should reconsider if other compact re-parametrizations of $\fr$, for example the one used in \citet{Terukina:2013eqa} would be physically more meaningful.

The reported bounds in \cref{tab:rel-errors-fR0-HS6} of the optimistic HS6 scenario, correspond to determinations of $\fr$ that are  $\fr=(5.0^{+ 2.2}_{-1.5} \times 10^{-6})$ with spectroscopic \GCsp\ alone,
$\fr=(5.0^{+ 0.91}_{-0.77} \times 10^{-6})$ combining WL, \GCph, and \XCph\ and $\fr=(5.0^{+ 0.62}_{-0.55} \times 10^{-6})$  with the full  \Euclid combination \GCsp+WL+\GCph+\XCph.
For the fiducial model HS5, $\fr=5\times10^{-5}$, which is already at the margins of the possible observational constraints, we have obtained constraints of the same order. This hints at the fact that \Euclid should be able to distinguish these two models from each other once data becomes available. 
For the model HS7 we find that the bounds get significantly less constraining, since it is a model very close to \lcdm\ with a similar growth of perturbations as compared to the standard cosmological model. Nevertheless we also find that under the optimistic scenario and using the full combination of \Euclid photometric and spectroscopic probes, this model can as well be distinguished from HS6 and from a \lcdm\ cosmology.

This highlights the significant amount of information that can be extracted from cross-correlating photometric probes, and by including a larger range in non-linear scales. In conclusion, \Euclid will be able to provide outstanding constraints on extensions beyond the concordance model thanks to its ability to probe the non-linear scales, where most of the information is contained. However, a good knowledge of the modelling of our theoretical observables at these scales, like the one used in this analysis, and a good control of the systematic uncertainties is required in order to reach reliable final results when data will be available.

\begin{acknowledgements}
%%%%
We thank Bill Wright for useful comments on the degeneracy between $f(R)$ models and neutrino masses. 
S.\ Cas. thanks  Sabarish V. M. and Sefa Pamuk for improvements in the development of the code \texttt{CosmicFish} used in this work.
S.\ Cas. acknowledges support from CNRS and `Centre National d'\'Etudes Spatiales' (CNES) grants during the early stages of this work. DS acknowledges financial support from the Fondecyt Regular project number 1200171.
IT acknowledges support from the Spanish Ministry of Science, Innovation and Universities through grant ESP2017-89838, and the H2020 programme of the European Commission through grant 776247. S.\ Cam.\ acknowledges support from the `Departments of Excellence 2018-2022' Grant (L.\ 232/2016) awarded by the Italian Ministry of Education, University and Research (\textsc{miur}), as well as support by \textsc{miur} Rita Levi Montalcini project `\textsc{prometheus} -- Probing and Relating Observables with Multi-wavelength Experiments To Help Enlightening the Universe's Structure' for the early stages of this project. P.\ N.\ is financially supported by `Centre National d'\'Etudes Spatiales' (CNES). The research of N.\ F.\ is supported by the Italian Ministry of University and Research (\textsc{miur}) through the Rita Levi Montalcini project ``Tests of gravity on cosmic scales" with reference PGR19ILFGP. N.\ F.\ and F.\ P.\  acknowledges the FCT project with ref. number PTDC/FIS-AST/0054/2021 
F.\ P.\ acknowledges support from the INFN grant InDark and the Departments of Excellence grant L.232/2016 of the Italian Ministry of Education, University and Research (\textsc{miur}).
K.\ K.\ is supported by the UK STFC grant ST/S000550/1, and the European Research Council under the European Union's Horizon 2020 programme (grant agreement No.\ 646702 ``CosTesGrav''). V.\ Y.\ acknowledges funding from the European Research Council (ERC) under the European Union's Horizon 2020 research and innovation programme (grant agreement No.\ 769130). L.\ L.\ was supported by a Swiss National Science Foundation (SNSF) Professorship grant (No.\ 170547).
A.\ P.\ is a UK Research and Innovation Future Leaders Fellow [grant MR/S016066/2].
We acknowledge open libraries support \texttt{IPython} \citep{4160251}, \texttt{Matplotlib} \citep{Hunter:2007}, \texttt{Numpy} \citep{Walt:2011:NAS:1957373.1957466}, and \texttt{SciPy} \citep{2019arXiv190710121V}.

\AckEC

\end{acknowledgements}

%\FloatBarrier

\bibliographystyle{aa}
\bibliography{biblio}

\appendix

\section{Further results}
\begin{figure}[htbp]
	\centering
	\includegraphics[width=0.95\linewidth]{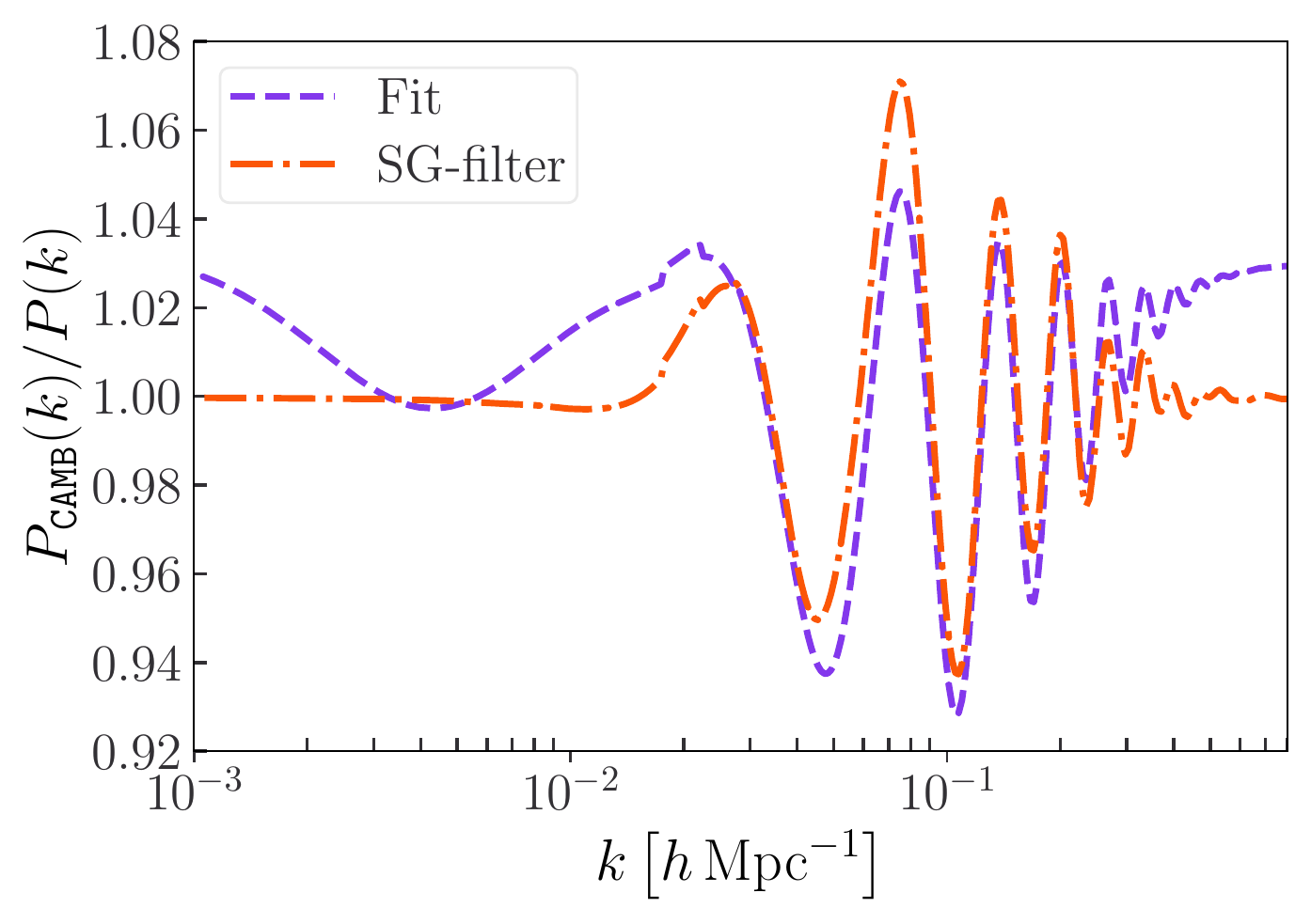}
	\caption{Comparison between the Eisenstein-Hu fitting method (in purple) and the Savitzky-Golay filter (in orange) for calculating the non-wiggle matter power spectrum. The curves show the ratio of the total linear matter power spectrum from CAMB divided by the corresponding smooth non-wiggle power spectrum.}
	\label{fig:SGfilter-A1}
\end{figure}

In this Appendix we show further results to highlight different aspects of our analysis.
In \cref{fig:SGfilter-A1} we show the difference between our determination of the non-wiggle matter power spectrum using the Savitzky-Golay (SG) filter (in blue) as compared to the standard Eisenstein-Hu method (in red) used previously in \citetalias{Blanchard:2019oqi}. As discussed in \cref{sec:spect} the SG method performs well and it is easily generalizable to modified gravity theories and models in which there is a scale-dependent growth of perturbations.

\begin{figure*}[htbp]
	\centering
	\includegraphics[width=0.45\linewidth]{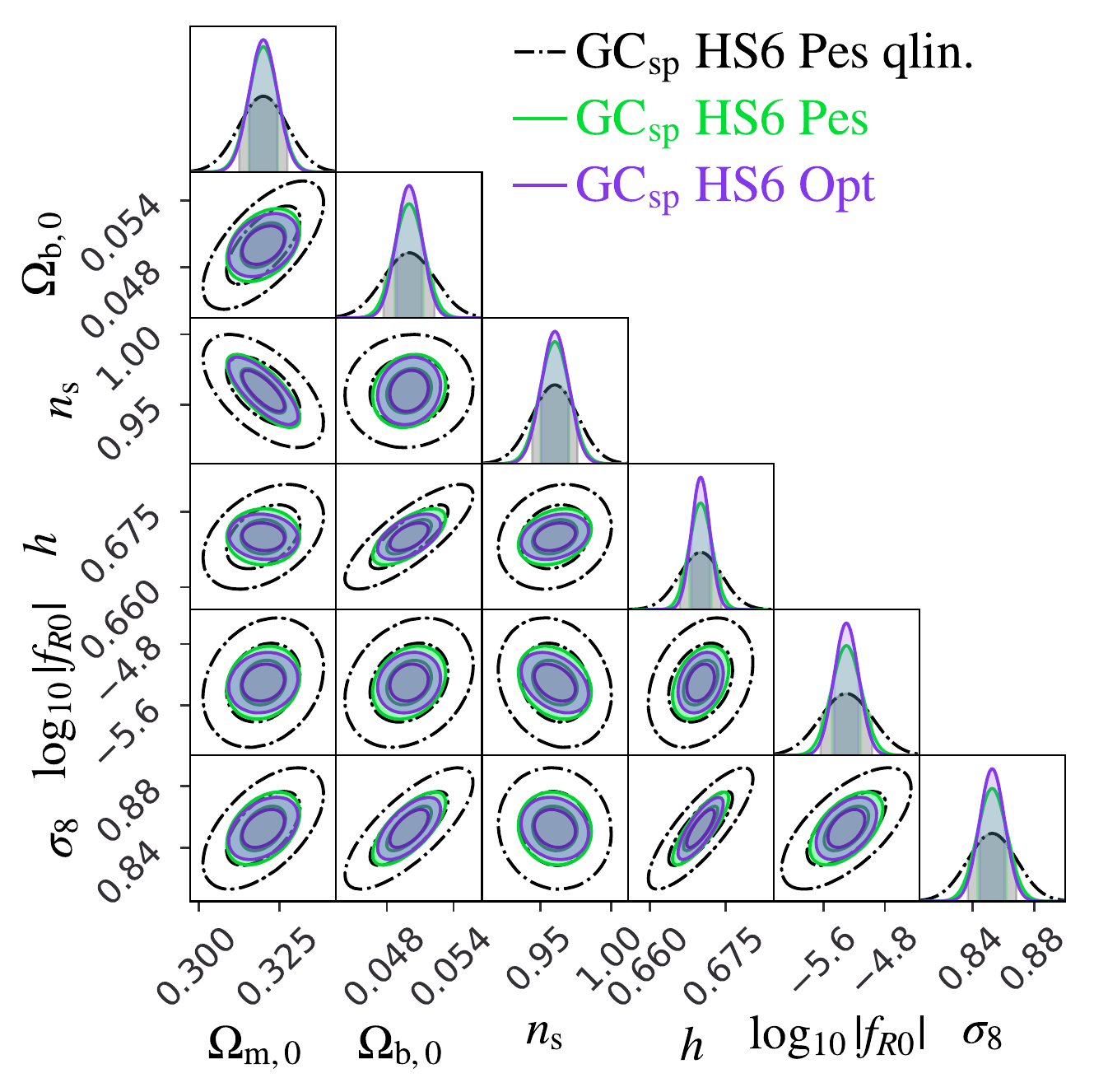}
    \includegraphics[width=0.45\linewidth]{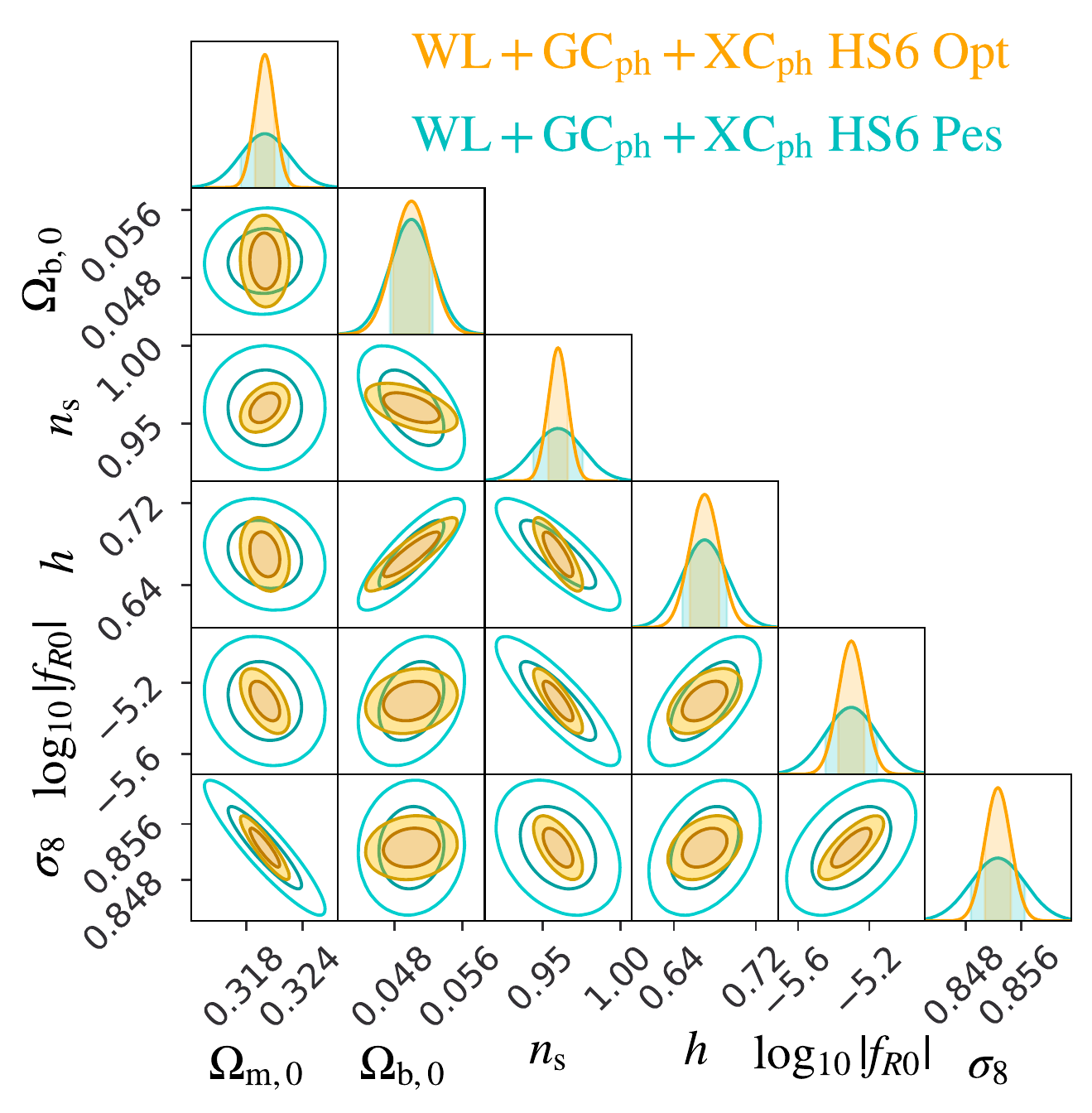}
	\caption{\textbf{Left panel:} 1 and 2$\sigma$ joint marginal error contours 
	on the cosmological parameters for a flat $f(R)$ model
		with $\fr=5\times10^{-6}$ (HS6) in the optimistic case for the GC spectroscopic probe.
		In black dot-dashed lines the quasi-linear case ($k_{\rm max} = 0.15$),
		in solid green the pessimistic case  ($k_{\rm max} = 0.25$) and
		in purple the optimistic case ($k_{\rm max} = 0.3$).
         \textbf{Right panel:} The full combination \GCsp+WL+\GCph+\XCph for the  HS6 case, 
		 comparing the optimistic (yellow, solid) and pessimistic (cyan, empty) specifications. 
		 See \cref{tab:specifications-ec-survey} for details on these scenarios.}
	\label{fig:ellipses-A2}
\end{figure*}

In \cref{fig:ellipses-A2} we show the comparison between the optimistic and pessimistic scenarios of the \Euclid probes as described in \cref{tab:specifications-ec-survey}, for the HS6 model. 
In the left panel we show for the spectroscopic GC probe the pessimistic quasi-linear (dotted-dashed lines),
the pessimistic (green) and the optimistic (purple) cases. One can see that, as expected, the difference between the pessimistic scale-cut at $0.25\,h\,{\rm Mpc}^{-1}$ and the optimistic one at $0.3\,h\,{\rm Mpc}^{-1}$, is not that visible in the final contours. However, the quasi-linear case ($ k_{\rm max} = 0.15\,h\,{\rm Mpc}^{-1}$), does discard important information in the quasi-linear regime and this has a major impact on the contours, degrading them by a factor 2 or more.
In the right panel of \cref{fig:ellipses-A2} we show the pessimistic (in cyan, empty contours) and optimistic (in orange, filled contours) scenarios for the full combination of photometric probes.
As discussed in \cref{sec:results} and shown in \cref{tab:rel-errors-fR0-HS6} the pessimistic errors on $\logfr$ are about a factor 2 to 3 larger than in the optimistic scenario, mainly due to the fact that non-linear scales contain crucial information on the scale-dependent growth of perturbations and the additional clustering caused by the fifth force present in $f(R)$. This highlights the importance in the future for real-data analysis to model these scales correctly and accurately to avoid introducing systematic errors in the analysis that could severely bias the parameter determination.

\begin{figure*}[htbp]
	\centering
	\includegraphics[width=0.45\linewidth]{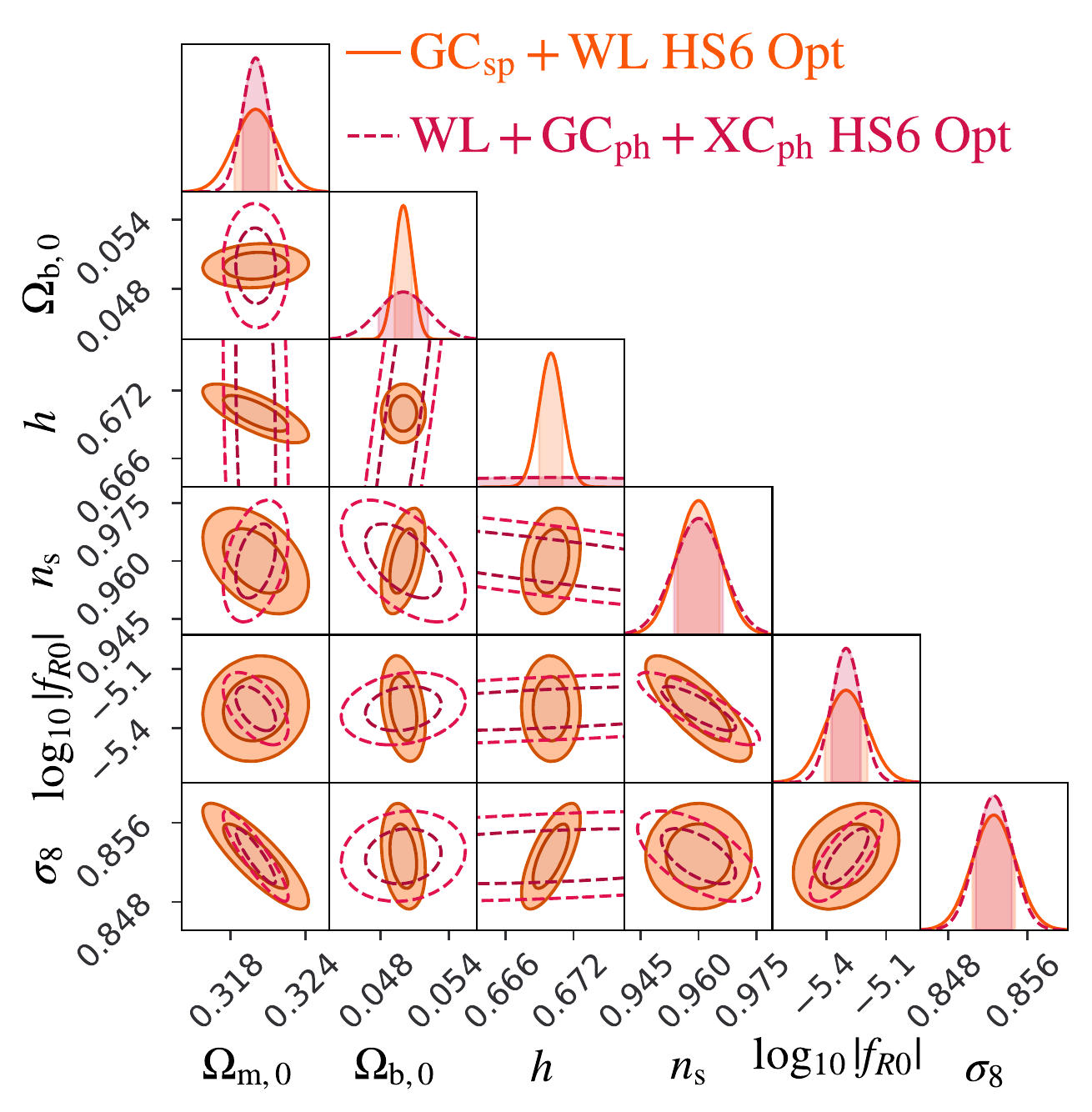}
    \includegraphics[width=0.5\linewidth]{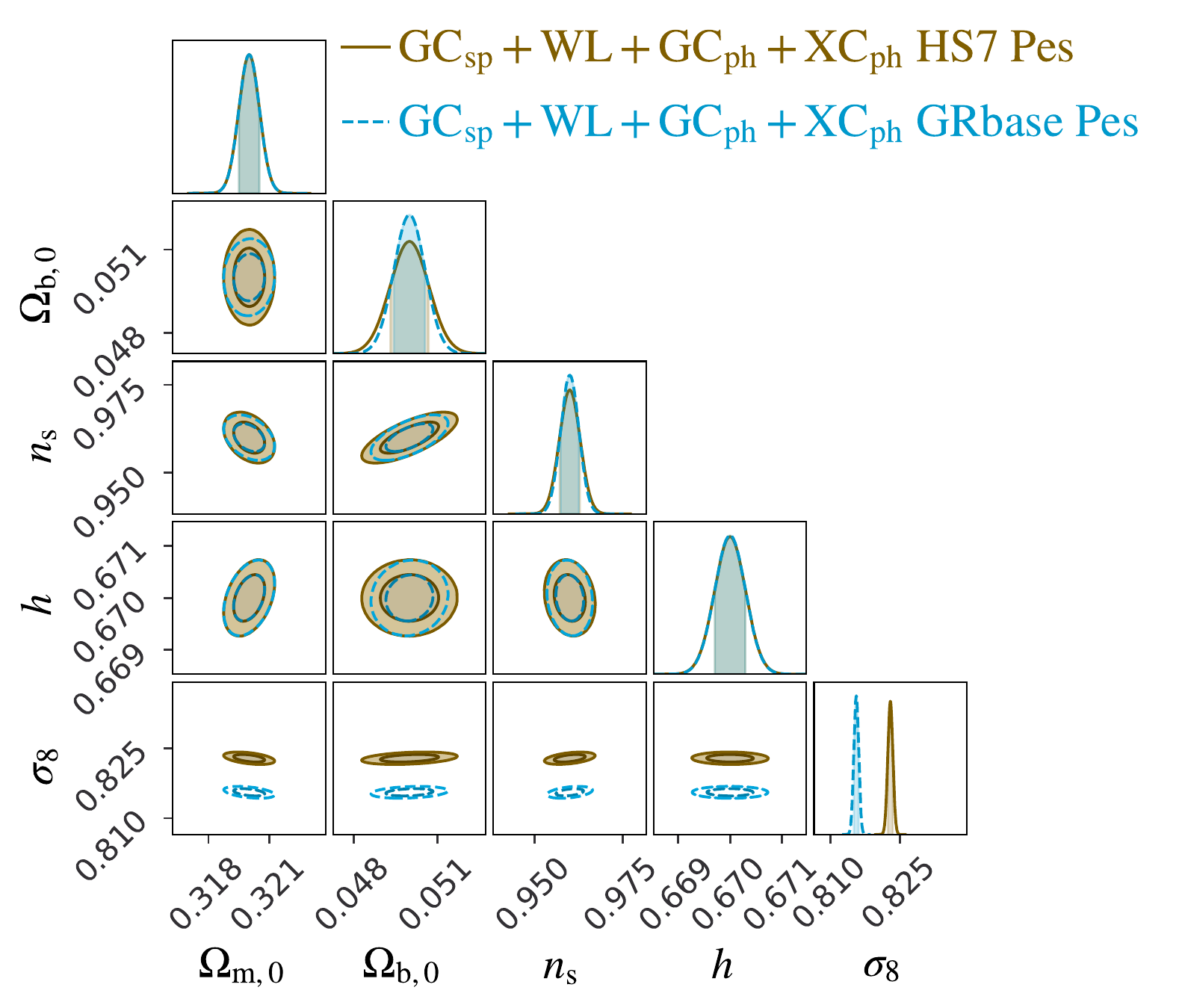}
	\caption{\textbf{Left panel:} 1 and 2$\sigma$ joint marginal error contours on the cosmological parameters for a flat $f(R)$ model
		with $\fr=5\times10^{-6}$ (HS6) in the optimistic case. 
		In orange the combination of the spectroscopic probe \GCsp\ and cosmic shear WL (same color code as in 
		\cref{fig:barplot-hs6-opt}), in red the photometric probes with their cross-correlation (WL+\GCph+\XCph).
		As explained in the main text, it is only the spectroscopic probe that is capable to break degeneracies with 
		the Hubble parameter $h$ and $\Omegab$, 
		such that the combination of spectroscopic and photometric yields indeed much better constraints 
		on the model parameter $\logfr$.
		The full combination of the spectroscopic and all the photometric probes including its cross-correlation is shown in \cref{fig:ellipses-hs6-opt}.
         \textbf{Right panel:} 
		 Comparison of the posterior contours when fixing (i.e. maximizing) the $\logfr$ parameter
		 from the Fisher matrices for the full combination \GCsp+WL+\GCph+\XCph\ for the HS7 case.
		 In dark yellow the HS7 case and in cyan dashed lines the GR baseline case. The contour shapes
		 match very well in all parameter subspaces, considering that in $\sigma_8$ their fiducial
		 values are different. This is expected from the fact that the HS7 model is very close to $\lcdm$.}
	\label{fig:ellipses-A3}
\end{figure*}

To study the effect of the probe combinations in more detail, we show in the left panel of \cref{fig:ellipses-A3}, for the $\fr=5\times10^{-6}$ model (HS6) and in the optimistic case, the main combinations possible with \Euclid primary probes.
Using the same color convention of the paper, we show in orange the elliptical 1$\sigma$ error contours for \GCsp+WL only, and in red the effect of the full combination of photometric probes (including its cross-correlation, $\XCph$). This is under the assumption that we can neglect the cross-correlation between the spectroscopic and photometric probes from the \Euclid observations, which has been shown to be a good assumption for Stage-IV surveys (see \cite{Taylor:2022rgy} and Paganin 2023, in prep.). For the $h$ and the $\Omegab$ parameter, as discussed in \cref{sec:results}, $\GCsp$ is the probe that gives the tighter constraints and the breaking of degeneracies in these parameter combinations, and it is what yields the improved constraints on $\logfr$ when combining \GCsp\ with the photometric probes.

Finally, in the right panel of \cref{fig:ellipses-A3} we show in dark yellow the fully marginalized contours for the HS7 model and in dashed cyan the contours for a \lcdm\ equivalent, after fixing (maximizing) the extra parameter $\logfr$. We do this to prove that the HS7 model is close enough to $\lcdm$ and that we recover very similar constraints when comparing the same number of free parameters, even if the HS7 model has a larger $\sigma_8$.  
In \cref{tab:rel-errors-fR0-HS5-HS7} the full list of $1\sigma$ marginalized errors can be found for HS5, HS7 and the \lcdm\ counterpart.

\end{document}